\documentclass[10pt, sigconf]{acmart}

\acmConference[MCPA - TechReport'18]{}{}{}
\setcopyright{none}
\renewcommand\footnotetextcopyrightpermission[1]{}
\usepackage{booktabs} % For formal tables
\usepackage{pbox}
\usepackage{subfigure}
\usepackage{epsfig,endnotes}
\usepackage{epstopdf}
\usepackage{adjustbox}
\usepackage{xspace}
\usepackage{xcolor}
\usepackage{hhline}
\usepackage{url}
\usepackage{comment}

\usepackage{cleveref}

\crefformat{section}{\S#2#1#3} % see manual of cleveref, section 8.2.1
\crefformat{subsection}{\S#2#1#3}
\crefformat{subsubsection}{\S#2#1#3}

\newif\ifcomment
\commenttrue
\newif\ifwatermark
\watermarkfalse

\ifwatermark
      \usepackage{draftwatermark}
      \SetWatermarkScale{.7}
      \SetWatermarkText{\shortstack{In Submission:\\Do Not Distribute}}
      \SetWatermarkColor[rgb]{1,0.88,0.88}
\fi

\ifcomment

    \newcounter{DPNumberOfComments}
    \stepcounter{DPNumberOfComments}
    \newcommand{\dpnote}[1]{\textcolor{orange}{\small \bf [DP\#\arabic{DPNumberOfComments}\stepcounter{DPNumberOfComments}: #1]}}

    \newcounter{AFNumberOfComments}
    \stepcounter{AFNumberOfComments}
    \newcommand{\afnote}[1]{\textcolor{magenta}{\small \bf [AF: \#\arabic{AFNumberOfComments}\stepcounter{AFNumberOfComments}: #1]}}

    \newcounter{GTNumberOfComments}
    \stepcounter{GTNumberOfComments}
    \newcommand{\gtnote}[1]{\textcolor{blue}{\small \bf [GT: \#\arabic{GTNumberOfComments}\stepcounter{GTNumberOfComments}: #1]}}

    \newcommand{\NOTE}[1]
    {
      {\footnotesize\it
        \begin{center}
          \begin{tabular}{|c|}
           \hline
            \parbox{0.85\columnwidth}{
              \medskip
              #1
              \medskip} \\
            \hline
          \end{tabular}
        \end{center}
        }
    }
\else
    \newcommand\dpnote[1]{}
     \newcommand\afnote[1]{}
      \newcommand\gtnote[1]{}
    \newcommand\NOTE[1]{}
\fi

\newcommand{\eg}{{e.g.,}\xspace}
\newcommand{\ie}{{\it i.e.,}\xspace}
\newcommand{\tool}{{\sf{MCPA}}\xspace}
\newcommand{\appclick}{{\sf{\mbox{app-click}}}\xspace}
\newcommand{\appstartup}{{\sf{\mbox{app-startup}}}\xspace}
\newcommand{\topalexa}{{\sf{\mbox{alexa-T100}}}\xspace}

\newcommand{\gradient}{$\nabla_b$\xspace}
\newcommand{\deltat}{$\alpha_t$\xspace}
\newcommand{\deltab}{$\alpha_b$\xspace}
\usepackage[ruled]{algorithm2e} % For algorithms

\SetAlFnt{\small}
\SetAlCapFnt{\small}
\SetAlCapNameFnt{\small}
\SetAlCapHSkip{0pt}
\IncMargin{-\parindent}

\makeatletter
\def\@copyrightspace{\relax}
\makeatother

\begin{document}

% Don't want date printed
%\date{}

\author{\Large Gioacchino$\,$Tangari,$\,$Alessandro$\,$Finamore$^\dagger\!,\,$Diego$\,$Perino$^\star,$} 
\affiliation{
\institution{
    \vspace{-1em}
    University College London,
    $^\dagger$O2 - Telef\'onica UK Ltd.,
    $^\star$Telef\'onica Research\\
    {\small gioacchino.tangari.14@ucl.ac.uk, alessandro.finamore1@telefonica.com, diego.perino@telefonica.com}\\
}
}

\title{Generalizing Critical Path Analysis on Mobile Traffic}

\begin{abstract}

Critical Path Analysis (CPA) studies the delivery of webpages to identify page resources, their interrelations, as well as their impact on the page loading latency. 
Despite CPA being a generic methodology, its mechanisms have been applied only to browsers and web traffic, but those do not directly apply to study generic mobile apps. 
Likewise, web browsing represents only a small fraction of the overall mobile traffic. 
%\afnote{The first sentence center CPA as a web-related thing, while instead is generic. I would replace it with ---Critical Path Analysis (CPA) studies how resources delivery and their interrelations impact a defined deadline. For instance, it is commonly used to study web pages rendering with respect the well known page load time (PLT). Unfortunately, currently available web-based CPA methodologies do not apply to generic mobile apps traffic, nor browsing represents a large share of mobile traffic.} \dpnote{I would keep it this way.}
In this paper, we take a first step towards filling this gap by exploring how CPA can be performed for generic mobile applications. We propose Mobile Critical Path Analysis (\tool), a methodology based on passive and active network measurements that is applicable to a broad set of apps to expose a fine-grained view of their traffic dynamics.
We validate \tool on popular apps across different categories and usage scenarios. We show that \tool can identify user interactions with mobile apps only based on traffic monitoring, and the relevant network activities that are bottlenecks. Overall, we observe that apps spend 60\% of time and 84\% of bytes on critical traffic on average, corresponding to +22\% time and +13\% bytes than what observed for browsing.
%studies the delivery of webpages to identify page resources, their interrelations, 

\end{abstract}
\maketitle

\section{Introduction}
\label{sec:introduction}
%Since its early days, the Web has been at the foundation of the Internet, and web browsing its most popular service. % a plethora of services and applications. 
Web browsing has been at the core of Internet services since its early days. 
%Numerous studies and articles report on how the quality of experience perceived by users (\eg Page Load Time (PLT)) is strongly related to the business values of both service and content providers. For instance, Amazon reported an increased revenue of 1\% for every 100 ms PLT reduction, or Mozilla reduced of 2.2s the PLT of its landing page increasing download conversions by 15.4\%\footnote{\url{https://webperformanceguru.wordpress.com/2010/07/}}. \dpnote{Should we add something about ISPs?}
Significant attention has been devoted to define metrics~\cite{bocchi2016,info3,navigation-timing,aft} and methodologies~\cite{wprofNSDI,aaltodoc,info3,goelPAM16} to unveil web pages content delivery dynamics, and systems to optimize content delivery~\cite{wang2016,webgaze2017,webprophet,klotski}. These efforts are justified to improve end-users quality of experience (QoE), while service providers are incentivized to optimize their systems as their revenues are linked to users QoE~\cite{fastcompany}.

However, web browsing is not at the center of user activities on mobile devices anymore.
Recent reports~\cite{smartinsights,flurrymobile} show that users spend less than 10\% of their time browsing, and more than 35\% on apps different than Facebook, streaming, gaming, and instant messaging.
%In other words, users do not use the browsers to consume the majority of the content they are interested into.
Such a trend is challenging also ads platforms where browsing on mobile devices generates half the conversion rate than desktop~\cite{grafik,monetate}. 

This progressive change in user interests and usage patterns is creating a \emph{gap in the literature}. State of the art metrics and methodologies have been forged in the context of web browsing, but they do not necessarily apply to generic mobile traffic. This is due to two main factors. First, there is the need to define a \emph{delivery deadline} capturing how long it takes to obtain some content. The most popular example is the page load time (PLT), which measures the time elapsed between a user clicking a URL and the browser firing the \texttt{onLoad} event indicating that the page has been loaded. Given the definition, such metric exists only for browsers. 
%Furthermore, despite W3C effort to standardize events related to web page download~\cite{navigation-timing,navigation-timing-level2}, different browsers might have different implementations, and measurements might not reflect dynamic content loading. 
A more generalized delivery deadline is the Speed Index (SI), which measures the average time at which the visible parts of the page are displayed~\cite{speedindex}. By focusing on the rendering process, SI is generic enough to be applied to services other than browsing, but it is an invasive technique as requires video screen recording, so it is not suitable to be deployed at-scale. Overall, the literature offers a flourished set of metrics (Yslow, Object Index, DOMLoad, etc.)~\cite{bocchi2016} but they all suffer from the lack of generalization or instrumentation complexity. Hence, the first challenge we identify is \emph{how can we define a delivery deadline that is generic enough and applicable to monitor generic mobile traffic?} %: at-scale for thousands of users?}.

Second, web pages structure is commonly leveraged to investigate content delivery performance. For instance, \emph{critical path analysis (CPA)} aims to identify which objects download impact a defined delivery deadline (\eg PLT), hence unveiling possible bottlenecks~\cite{wprofNSDI,lighthouse,klotski}.
This analysis is possible because web page object relationships are easy to extract (\eg inspecting source files, or the Document Object Model - DOM).
%\ie the study of web pages to identify objects inter-dependencies, and how they impact a web page loading time. By studying the web page structure and how is rendered by the browser~\cite{wprofNSDI}, and by defining a delivery deadline (\eg Page Load Time -- PLT, DOMLoad, Above The Fold -- AFT time) it is possible to dissect the dynamics of a web page download~\cite{navigation-timing,navigation-timing-level2}. This analysis allows providers and developers to identifies bottlenecks in the delivery of webpages (\eg JS execution, network activities) that could be optimized to improve user QoE.
Unfortunately, mobile content is not necessarily delivered in the form of a web page. Even if dependencies between objects are expected to be present, their identification is not trivial. %We believe that rationale behind CPA still holds, but 
Hence, the second challenge is \emph{how can we identify which flows carry critical content for QoE, and how they relate to each other?} % And, can we identify the relationships between different flows when considering generic mobile traffic?}

In this paper we present \tool, a methodology that generalises CPA for mobile traffic. \tool brings fine-grained visibility into any mobile app traffic, and further highlights which components are critical. To do so, the traffic is processed in three phases.
First, the traffic aggregate is split into \emph{activity windows}, each (possibly) corresponding to different user interactions with an app. Second, within the activity windows, a download waterfall is constructed to capture traffic dynamics over time, different metrics related to L4 and L7 dynamics are collected, and a delivery deadline is established.
Finally, within each waterfall we identify which activities impact performance.

We validate our methodology on 18 popular apps and web browsing as well, generating traffic from an instrumented Android phone. We show that using a purely based traffic monitoring methodology, \tool is sufficient to capture fine-grained traffic dynamics. Specifically, we can split aggregate traffic into windows each associated to a different user action with more than 84\% accuracy (\cref{sec:intervals}).  We define two traffic metrics based on monitoring the volume of bytes exchanged, and show that they well resemble the more complex state of the art AFT and SI (\cref{sec:waterfall}). Finally, we perform CPA by mean of active experiments. When considering browsing, \tool output is a superset of the critical traffic identified by state of the art Google Lighthouse~\cite{lighthouse}.  As for mobile applications, the time spent on the critical path is 55\% in average,  significantly larger than browsing where the time spent on critical path is 38\%. We  observe this time is mostly related to application control logic. \tool source code and experimental datasets are publicly available.\footnote{\small\tt\mbox{\url{https://github.com/finale80/mcpa}}}

\section{Related Work and \tool challenges}
\label{sec:related}

Mobile traffic has mostly been studied at an aggregated level (per-connection latency, throughput, etc.)~\cite{falaki2010first,antmonitor,fioreSURVEY,bustamanteIMC14}, or focusing on specific protocols (e.g., DNS~\cite{almeida2017}, SPDY~\cite{erman_2013CONEXT}, MPTCP~\cite{han2015,han2016}). Exceptionally, a few studies take a step further. For instance, Panappticon~\cite{panappticonISSS13} and AppInsight~\cite{appinsightOSDI12} enable fine-grained view on users engagement with apps by respectively tapping into Android components and studying app binary files; QoE Doctor~\cite{qoedoctorIMC14} focuses on performance issues (\eg high latency) by measuring radio resource allocation and user interface interactions; %, all without accessing apps or OS source code; 
Prometheus~\cite{prometheusHOTMOBILE14} tries to bridge network metrics with user experience via machine learning.

Despite their merits, these tools focus on system information (\eg radio resources, operating system calls, multi-threading) rather than digging into the role of content download and network protocols dynamics. Conversely, studies focusing on web traffic, despite being limited to this traffic class only, represent the state of the art regarding how to dissect traffic dynamics. In the remainder of this section we review this literature, and we highlight the challenges in applying currently available methodologies to study generic mobile traffic.

%There is an abundant literature focusing on web browsing performance. 
%This includes studies based on  \emph{objective metrics}, i.e., measurements gathered passively monitoring content delivery~\cite{falaki2010first,flywheelNSDI15,erman_2013CONEXT,qian2014,ma2015},  studies incorporating \emph{subjective metrics}, i.e., measurements obtained via direct feedback from end-users~\cite{eyeorg,bocchi2016,gao2017,kelton2017}, as well as work focusing on critical path analysis.  In this work we focus on objective performance, and on critical path analysis: in the following we therefore review work related to these two research areas.

%\subsection{Objective Performance Metrics} 
\subsection{Performance metrics and delivery deadlines} 

Beside generic metrics such as latency and throughput, most of the metrics in literature are defined in the context of web traffic. We can split those into two classes: \emph{objective metrics} are delivery deadlines quantifying the time needed to obtain some content~\cite{flywheelNSDI15,erman_2013CONEXT,falaki2010first,ma2015,qian2014}; \emph{subjective metrics} are defined considering direct feedback from end-users (\eg mean opinion score - MOS)
and can include factors beyond content delivery~\cite{bocchi2016,gao2017,webgaze2017,eyeorg}.
For the purpose of this work, we focus only on objective metrics, which we can further split into \emph{time instant} and \emph{time integral} metrics.
%Objective performance metrics can be divided in two classes: \emph{Time instants} and \emph{Time integrals}. 

%\vspace{2pt}
\noindent\textbf{Time instant} metrics capture specific instants across the whole events timeline of the content delivery. 
The most accurate instant metric is Google's AFT which measures the time at which the content shown in the visible part of a webpage is completely rendered~\cite{aft}. This definition is not web traffic specific, although the metric has been applied only to browsing traffic. AFT computation requires a video screen capture, and accurate video post-processing as the presence of dynamic elements, such as animations and roll ads, can introduce biases~\cite{eyeorg}. These costs limit the use of AFT for small scale studies on instrumented devices. A recent work shows that AFT could be approximated leveraging information about objects position in a webpage, but this technique is complex to be applied outside browsers~\cite{aaltodoc}. Despite being less accurate, PLT is the most widely adopted metric. %It refers to the time at which the browser fires the \texttt{onLoad} event. 
Other known deadlines are the \emph{Time To First Byte} (TTFB), the \emph{Time To First Pixel} (TTFP), the time at which the parsing of the \emph{Document Object Model} (DOM) is completed. W3C has also defined the navigation timing guidelines~\cite{navigation-timing-level2}, a series of specific events happening during a webpage rendering, but their implementation may differ across browsers.
%Those metrics cannot be applied to quantify the quality of experience of mobile applications as they are defined for webpages and based on events generated by browsers that are not available in mobile apps.

%\vspace{2pt}
\noindent\textbf{Time integral} metrics capture the cumulative effect of events until a specific point in the timeline is reached. %The goal  is to capture the evolution over time of a given type of events rather than a single timing information. 
The most popular example is Google's SI~\cite{speedindex}, %which measures the average time at which the visible parts of the page are displayed~\cite{speedindex}. 
%More in detail, SI 
which is obtained by integrating over time the residual rendering left to reach the AFT. Given the definition, SI suffers from the same limitations AFT does.
%Other integral metrics have been proposed following the SI definition. For instance, 
ObjectIndex and ByteIndex are two alternative integral metrics that respectively capture the evolution of objects and bytes delivery until the PLT~\cite{bocchi2016}. %However, being web browsing specific, they cannot be applied to generic mobile traffic. %We refer the interested readers to~\cite{bocchi2016} for a more in-depth analysis of both time instant and time integral metrics, and their relationships.

%\vspace{2pt}

\noindent\textbf{Challenges:} Metrics like PLT, which are based on internal application ``hooks'', cannot be applied to generic mobile apps as there are no standard APIs, neither at app nor at operating system level, to expose these information. %ed by both apps and operating systems. %across apps, and given that  operating systems do not offer a standard solution. 
Differently, we argue that \emph{AFT and SI are valid delivery deadline for generic mobile apps}, as they capture the actual screen rendering and do not depend on app internals (\cref{sec:overview}).
%\afnote{the 2 sentences before seem disconnected} 
However, their measurement cost is a barrier for their adoption. %, they present a barrier have limited applicability.
To enable at-scale measurement, a cheaper alternative is to opt for metrics based on \emph{passive traffic measurement} to compute either on-device (\eg via VPN solution which avoid rooting devices) or in-network (\eg monitoring middle-boxes are very common in mobile networks).
We are therefore interested in understanding \emph{what passive metrics are available, when they can be applied, and what bias they introduce with respect to AFT and SI.}
%\afnote{not sure this point is clear: are you saying that we are interested in understanding if AFT/SI are usable for mobile traffic too? I guess not because you said before that they are...or perhaps the point is understand if there is anything beyond AFT/SI that can be used (I think this is what I wanted to say)}

%\noindent\textbf{Challenges:} \change{To some extent, all the metrics discussed are ``invasive'' as they either require hooks within the browser, or extensive on-device instrumentation. Such rigidity is not suited to cope with a large amount of apps, so we need to define a technique easier to apply.
%One solution could be to opt for metrics based on \emph{passive traffic measurement} to be applied either on-device (\eg via VPN solution which avoid rooting devices) or in-network (\eg monitoring middle-boxes are very common in mobile networks).
%We acknowledge that AFT and SI are the best metrics to reflect rendering effects possibly hard to see just looking at the traffic.
%However, given their cost, we are interested in understanding \emph{what ``cheaper'' options are available, when do they apply, and how distorted they are}}.

\subsection{Critical Path Analysis - CPA\label{sec:related-critical}}
CPA allows to dissect traffic dynamics within the boundaries of a delivery deadline. It has been successfully applied to understand web traffic, but methodologies and terminology can vary.
To the best of our knowledge, the first tool leveraging CPA is WProf~\cite{wprofNSDI} (and its follow ups Shandian~\cite{shandianNSDI16}, and WProfX~\cite{wprofxWWW16}), a system that requires augmenting the browser with a profiling engine to capture the \emph{dependency graph} for any given webpage. Such graph structures the activities related to both rendering as well as content dependencies as visible in the webpage DOM. Given a graph, WProf defines the critical path as the longest path of activities such that reducing the duration of any activity not on the critical path does not impact the webpage PLT.

Recently, Google added Lighthouse~\cite{lighthouse} to the Chrome devtools suite to automate webpages auditing. 
Lighthouse offers a richer output than WProf, including different deadlines (First Meaningful Painting, First CPU idle, SpeedIndex, etc.), as well as a report on resources that can block the rendering. %, and a film strip showing the rendering evolution.
%Such interface is available in modern browser, but with Lighthouse the waterfall is studied to extract the critical component. 
%Differently from WProf, Lighthouse uses network priority (\ie how the browser decide what should be fetched first) as a proxy for identifying render-blocking critical resources.\footnote{https://developers.google.com/web/tools/lighthouse/audits/critical-request-chains} It follows that Lighthouse offers a reacher output than WProf, as multiple \emph{critical chains} (as named by the tool) can be associated with the same rendering-blocking stages of the load.
To some extent, Lighthouse output is an evolution of a webpage \emph{download waterfall}, \ie a gantt chart picturing the evolution of the network communications triggered during a webpage load. All modern browsers allow to dissect traffic dynamics via a waterfall, % is a key tool to study the critical path, and modern browser have builtin tools to display and interact with such charts. %For instance, Lighthouse is basically a tool extracting key components from a waterfall, \afnote{I think we can drop the bit about lighthouse} while 
and systems like KLOTSKI~\cite{klotski} further build on waterfalls to find activity patterns invariant to PLT performance. % across multiple waterfalls obtained from the same webpage. % Initiatives like \emph{webpagetest.org} and \emph{gtmetricx.com} are cheap options to get performance metrics quickly, but likely useful for CPA only for expert users.}

\noindent\textbf{Challenge:} All these tools have slightly different critical path definitions. They also heavily rely on ``hooks'' specific to browsers internals, so they are unappealing to study mobile apps. %As we cannot rely on the same methodologies, we need \emph{a novel definition of the critical path for generic mobile applications}. 
At the core of CPA there is the need to identify dependencies between activities, and this is particularly challenging to do only based on passive measurements.
Hence, we want to understand if \emph{active experiments}, such as traffic throttling, can complement passive measurements to create a more effective methodology to spot traffic impacting the delivery deadline. 

\section{\tool Overview}
\label{sec:overview}

In this section, we introduce \tool, our methodology to perform CPA on generic mobile apps. First, \tool identifies activity windows, \ie user interactions with apps. %, where CPA should be performed. 
Each activity window is profiled to extract network activities, measure the delivery deadline, and finally extract the critical traffic.

\noindent
\textbf{Activity windows (\cref{sec:intervals}).}
In the context of web traffic, CPA is performed for every webpage retrieval.
This includes all activities in response to directly typing a URL, refreshing or aborting the load of a webpage, clicking a link within a page, etc. For webpages, those activities can be easily identified using APIs provided by browsers. However, such mechanisms are not available to study generic mobile apps, so alternative approaches need to be considered.  
%Despite mobile devices are always connected, they generate ``burst'' of traffic~\cite{falaki2010first,Stober:2013:YSY}. \change{These bursts are associated either to background application activities or human interaction with applications. As background traffic does not have impact on user experience, it is relevant to perform CPA on the latter cases only. 
One option is to log user clicks, scrolls, currently displayed apps, and use such detailed information to partition the traffic based on user engagement. However, in an at-scale scenario, \ie without full control on the devices, logging actual user interactions is almost impossible. Another option available is to apply ``cheaper'' passive traffic analysis heuristics. In fact, mobile traffic is bursty in nature~\cite{falaki2010first,Stober:2013:YSY}, \ie the traffic presents \emph{activity windows} when the user is interacting with the phone, interleaved by ``idle'' periods. 
An optimal split associates a different user action to each window, but depending on traffic conditions and apps characteristics this might not always be possible. In \cref{sec:intervals} we discuss heuristics for partitioning the traffic based on passive measurements and we evaluate their accuracy.

\noindent
\textbf{Download waterfall and performance metrics (\cref{sec:waterfall}).}
For each activity window we need to define a set of metrics and identify the activities involved in the delivery of contents. CPA for webpages requires to instrument the browser to extract all activities participating to both the download and rendering tasks. However, to do the same for generic mobile apps %it is not possible to obtain fine-grained statistics about processing activities. Indeed, this 
would require to either reverse engineer every app, or to instrument their source code or the operative system~\cite{appinsightOSDI12,panappticonISSS13}. 
%This approach is clearly not viable considering the large amount of applications most of which are closed source. 
 %\dpnote{this sentence may create confusion. Indeed, we say we cover both controlled and in the wild scenario. But we never cover this CPU/GPU part.} \afnote{not sure what you mean with ``we cover both controlled and in the wild'', and I don't understand where you get lost. The text clearly says we DON'T capture rendering effects. To elaborate, in the previous paragraph we discuss what one can and cannot do to make the methodology ``portable''. The same should apply here. Since in the related work we discuss rendering related metrics, and since we use AFT/SI, we need to reiterate here on what this means. If you are on-device we can track those, as well as other resource usage, and WProf does that when creates the dependency graph. This is BESIDE capturing network metrics. The point is to reiterate this make the methodology rigid and we consider only network traffic to make things easier to use.} 
The approach of \tool is to focus only on network activities and to report per-flow metrics for both transport (TCP, UDP, QUIC) and application (DNS, HTTP, HTTPS/TLS, Facebook Zero - FB0) protocols. These activities are visually represented in the form of a download waterfall. %(\cref{sec:cpa}).
%identify activities and measure associated statistics. As we do not have access to application internals, the analysis should be limited to network activities; 
%we therefore need to construct a
%\tool  
%Beside bringing visibility on traffic dynamics, the waterfall analysis enables us to define a delivery deadline. 

Once the different activities are identified, a delivery deadline should be set to capture the quality of experience perceived by users.
In a fully controlled environment, the best available option is to apply AFT and SI (\cref{sec:related}). We argue they are still valid to study generic mobile traffic, but we are not aware of any work in the literature proving this. Indeed, the end of a user action on an app is generally marked by visual changes, and this applies to apps wrapping browser(-like) functionalities (\eg social, news, e-commerce), as well as to more interactive apps such as messaging ones (\eg the end of a message delivery triggers a check mark on screen). %, and those metrics steadily apply. 
%This is true also for interactive apps (\eg messaging) as every time a new content is sent or received it should be possible to define an activity window based traffic exchange patterns.  
However, both AFT and SI capture events related to rendering. % that are not possible to capture otherwise.
%\afnote{I think we should start with this, and them move to network metrics}
In an at-scale scenario screen recording is not possible, %and we need to define a deadline based on passive measurement.
%In other words, 
so rather than looking for exact estimates of user experience, we are interested in defining a proxy for AFT/SI, yet sufficient to identify critical activities, based on passive measurements.
%If screen recording is not possible, for instance in large scale or in network measurements, we need to define metrics purely based on passive observation.
In \cref{sec:waterfall} we discuss how \tool creates waterfalls, we introduce our delivery deadlines, and we compare them against AFT/SI.

\noindent
\textbf{Critical Path (\cref{sec:cpa}).} Finally, \tool identifies which activities of a waterfall constitute the critical path. To do so, we rely on active experiments, \ie we observe how the delivery deadline changes when throttling the traffic on a per-domain basis. %of the download waterfall and identify which ones have an impact on the TDT. 
In other words, %by intentionally slowing down some traffic, 
if a macroscopic delay is observed on the overall delivery when delaying some traffic, we can conclude that a domain, and the related traffic, is critical.
The same principle also applies to discover relationships among domains.
%Sec.~\ref{sec:cpa} details the CPA analysis along with the critical path characteristics we found for the apps we considered. 

\tool is built upon \texttt{pcap2har}, a Python open source tool transforming pcap files into webpages HAR files,\footnote{\url{https://github.com/andrewf/pcap2har}} which we modified and extended to handle generic mobile traffic (including TLS/HTTPS, QUIC, FB0).

\section{Dataset}
\label{sec:dataset}

\noindent{\bf Mobile Apps.} 
We select 18 popular apps across 7 categories: Social (Twitter, Facebook, Instagram), Messaging (WhatsApp, SnapChat, Messenger), News (CNN, BBC, Newsbreak), Geo-based (Google Maps, Uber), Shopping (Letgo, Amazon), E-mail (Microsoft Outlook, Gmail), and Streaming (Youtube, Spotify, Soundcloud).
We intentionally left out Games and Productivity apps as they are known to generate little network traffic, which is likely related to ads~\cite{almeidaWWW18}.
Conversely, we focus on very popular apps according to both vendors~\cite{sandvineGIP18}, and 3rd party\footnote{\url{https://www. androidrank.org}} rankings, to create a set of apps sufficiently diversified to assess if there is a case to use passive and active analysis to perform CPA.
We further consider web browsing by studying the top-100 Alexa websites (\topalexa).

\noindent{\bf Traffic Scenarios.}
We consider two traffic scenarios: \appstartup and \appclick.
The former considers the traffic generated in the first 60s after the app is launched.\footnote{This time is more than double the maximum startup time observed in our experiments.} In the latter,  relevant user interaction sequences are emulated based on common behaviors with the apps, such as select a video/song, a news, scroll an email, send a chat message, etc. To this end, we define ad-hoc patterns, each with multiple \texttt{input tap} events uniformly distributed within [0,10s]. For example, for the Letgo shopping app, the sequence is: search by category; show top results; select random item; show price and geographical location (all sequences listed in Table \ref{tab:usagepatterns}).
%using a randomized sequence of clicks generated via \texttt{monkey}\footnote{\url{https://developer.android.com/studio/test/monkey}} and \texttt{input tap}\footnote{\url{https://www.raizlabs.com/dev/2017/09/automating-input-events-android/}} Android utilities and uniformly distributed in [0,10s]. \afnote{In particular, the monkey is not performing app stress tests via random clicks across the screen, but rather uses ad-hoc patterns we created to resembling actual user actions, with a random inter-time between the events pattern.}
 %[0,2.5], [2.5,5], [5,7.5], and [7.5,10] seconds. For both scenarios the clicks are 
%Similarly to the choice of apps, the rational for the selected scenarios is to validate \tool with a large spectrum of traffic patterns.

\noindent{\bf Data collections.}
%Our experiments are performed on a Nexus 5 running Android 6.0.1, and using a SIM of a European mobile carrier.
%We instrumented the device to collect pcap files (via \texttt{tcpdump}) as well as the video screen record (via Android \texttt{screenrecord} utility\footnote{\url{https://developer.android.com/studio/command-line/adb}}). For \topalexa dataset, we also used Google Lighthouse and Chrome's devtools to extract performance indicators and critical path information. %\afnote{we need to say that the 4G connectivity is emulated, right?}
%We performed 10 experiments for each app and scenario, making sure that only the app under study was running.
%In regards to the video recording, as highlighted in~\cite{bocchi2016}, the computational complexity can bias the experiments, artificially slowing the rendering.
%However, we observed that this complexity does not impact any measured value making it sustainable for our evaluation (for details see \cite{mcpatechreport}).
Our experiments are performed on a Nexus 5 running Android 6.0.1, and using a SIM of a European mobile carrier.
For each app and scenario we ran 10 experiments, with the device instrumented to collect pcap files (via \texttt{tcpdump}) as well as the video screen record (via Android \texttt{screenrecord} utility\footnote{\url{https://developer.android.com/studio/command-line/adb}}). For \topalexa dataset, we also use WProfX, Google Lighthouse and Chrome's devtools to extract performance indicators and critical path information. %\afnote{we need to say that the 4G connectivity is emulated, right?}
%We performed 10 experiments for each app and scenario, making sure that only the app under study was running.
In regards to video recording, as shown in~\cite{bocchi2016} the additional computation can bias the experiments, artificially slowing the rendering.
We verified that this effect is not present in our results (\cref{sec:waterfall}).

%\begin{figure*}[t]
%\centering
%\includegraphics[width=0.8\textwidth]{fig/waterfall_new2-crop}
%\caption{Example of waterfall for Letgo.}
%\label{fig:waterfall-example}
%\end{figure*}

\section{Activity Windows}
\label{sec:intervals}

\begin{figure}[!t]
%\begin{minipage}{.47\textwidth}
  \centering
  \includegraphics[width=\columnwidth]{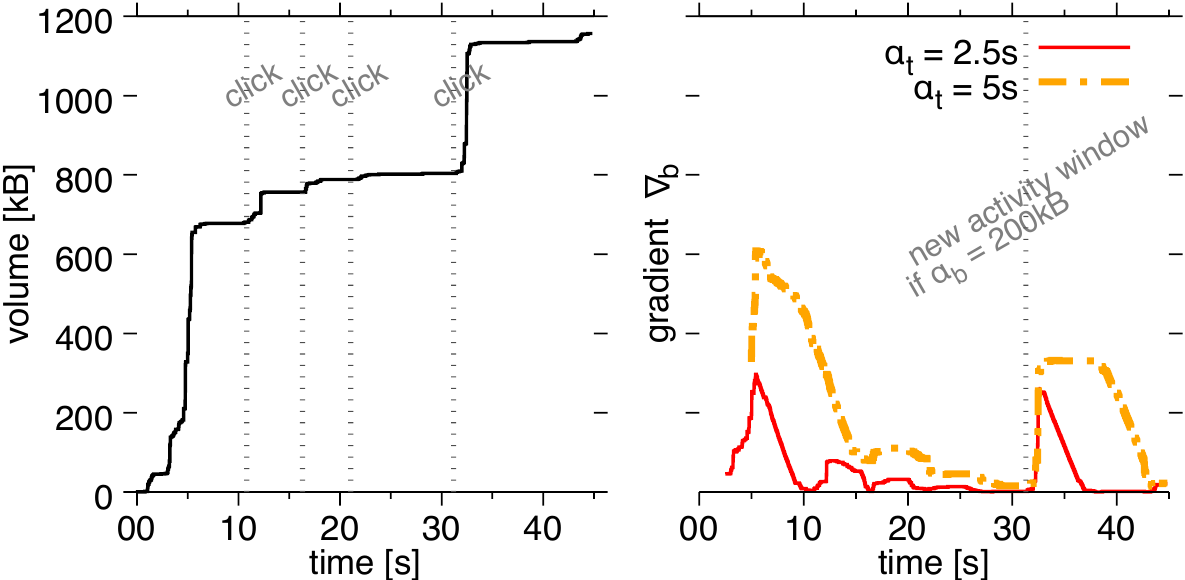}
  \caption{Activity windows: cumulative traffic when using the CNN app (left); traffic gradient \gradient (right).\label{fig:waterfall-splitting}}
\end{figure}

Mobile devices are constantly connected to the network, so they generate a continuous stream of connections. 
%Conversely, user engagement is occasional, so the connections stream have to be processed accordingly to identify \emph{activity windows} where users interact with the device. Such split is easy to do using device screen logs reporting on clicks, scrolls, etc., at the cost of running tests only on a limited set of properly instrumented phones. To overcome this limitation, passive measurements could be adopted to capture the bursty nature of mobile traffic. For instance, consider Fig~\ref{fig:waterfall-splitting}~(left) showing the cumulate traffic observed when a user interacts with the CNN app. Notice how volume abruptly increases in response to users actions. In this section we investigate if we can leverage these traffic bursts and traffic idle periods to identify activity windows. 
Conversely, user engagement is occasional, hence the connections stream has to be processed in order to identify those time intervals where users interact with the device. Ideally, the traffic stream should be split so that each partition corresponds to a relevant QoE-related user interaction. We call these partitions \emph{activity windows}. Such windows can be obtained using granular device-screen logs reporting on clicks, scrolls, etc., at the cost of running tests only on a limited set of instrumented phones. 

To enable large scale analysis built on network measurements, the same split should be performed by looking at traffic characteristics only. To this end, we can exploit the bursty nature of mobile traffic, where bursts of bytes are likely to correspond to user engagement with an app. For instance, consider Fig~\ref{fig:waterfall-splitting}(left) showing the cumulative traffic observed when a user interacts with the CNN app. Notice how volume abruptly increases in response to users actions. In this section we investigate how and to what extent traffic bursts and idle periods can be used to identify activity windows.

%except in the interval 0-5s corresponding to the app start up.}
%Overall, a sudden change in the traffic volume can be an indication of user engagement with the app, although it might not always be the case. The challenge is how to split the traffic to create one activity window for each click. 

\subsection{Partitioning policies.}
We consider two possible policies to partition the traffic generated by a mobile device.

\noindent \textbf{Na\"ive.} The first policy relies on a single threshold to identify ``long'' idle periods. That is, a connection is associated to a new window if its traffic starts after an idle period longer than \deltat, otherwise it belongs to the current window.

\noindent \textbf{Gradient.} A more refined policy creates a new window if a ``large'' burst happens after a ``long'' idle period. To do so, we combine two thresholds: \deltat and \deltab.
We use \deltat to define a sliding window where we monitor the \emph{gradient} \gradient of the volume. For instance, consider \deltat = 5s. All traffic in the first 5s is accumulated. Then, we progress the sliding window, accumulating the traffic entering, and removing the one falling outside the window. In this way \gradient has a positive slope when traffic is exchanged, and negative (or no) slope for idle times. Fig.~\ref{fig:waterfall-splitting}(right) reports \gradient for \deltat = 2.5s and \deltat = 5s. 
%Notice how smaller \deltat values create a more impulsive gradient. 
Using the gradient, we define a new activity window if we observe at least \deltab bytes exchanged after an idle period of \deltat.
%In other words, when \gradient is not positive, we set a timer and check if within the next \deltat the gradient increases of at least \deltab. 
%As such, \deltat is used for the sliding window as well as to identify idle periods.
For instance, considering \deltab = 200kB, in Fig.~\ref{fig:waterfall-splitting}(right) \gradient reaches the threshold at 5.2s and 32s. However, we identify an activity windows only at 32s as it is preceded by an idle larger than \deltat = 2.5s (no windows found for \deltat = 5s). %Notice also that the activity windows starts when the traffic start to grow. 

%user interactions with an application from a continuous stream of packets (\ie pcap file). The output of this first phase is a pcap file segmented in a sequence of activity intervals.  

%Based on these observations the rationale of our algorithm  is that a user interaction with an application can be detected on a network trace by observing a transition from an \emph{idle} period, where little data is exchanged, to an \emph{active} period of time where significant amount of bytes retrieved.  Active periods are caractherized as a continuous period of times where for every interval $T_B$ at least $\alpha B$ bytes have been exchanged. $B$ denotes the amount of Bytes exchanged on every experiment, and $\alpha$ and $T_B$ are input parameters. Note that $B$ would need to be tuned for \tool usage in the wild; this is out of the scope of this paper and left as future work. Idle periods are the interval of times for which the previous condition does not hold. Transition from idle to active periods are identified as user interactions with the application and the input file is segmented accordingly.

%\dpnote{we need this plot and what do we describe? Merge with text above in case}
%Fig.~\ref{fig:waterfall-splitting} illustrates how \tool identifies user activities (blue lines) and compares them to actual user actions (red lines). 

\begin{figure}
\centering
\includegraphics[width=\columnwidth]{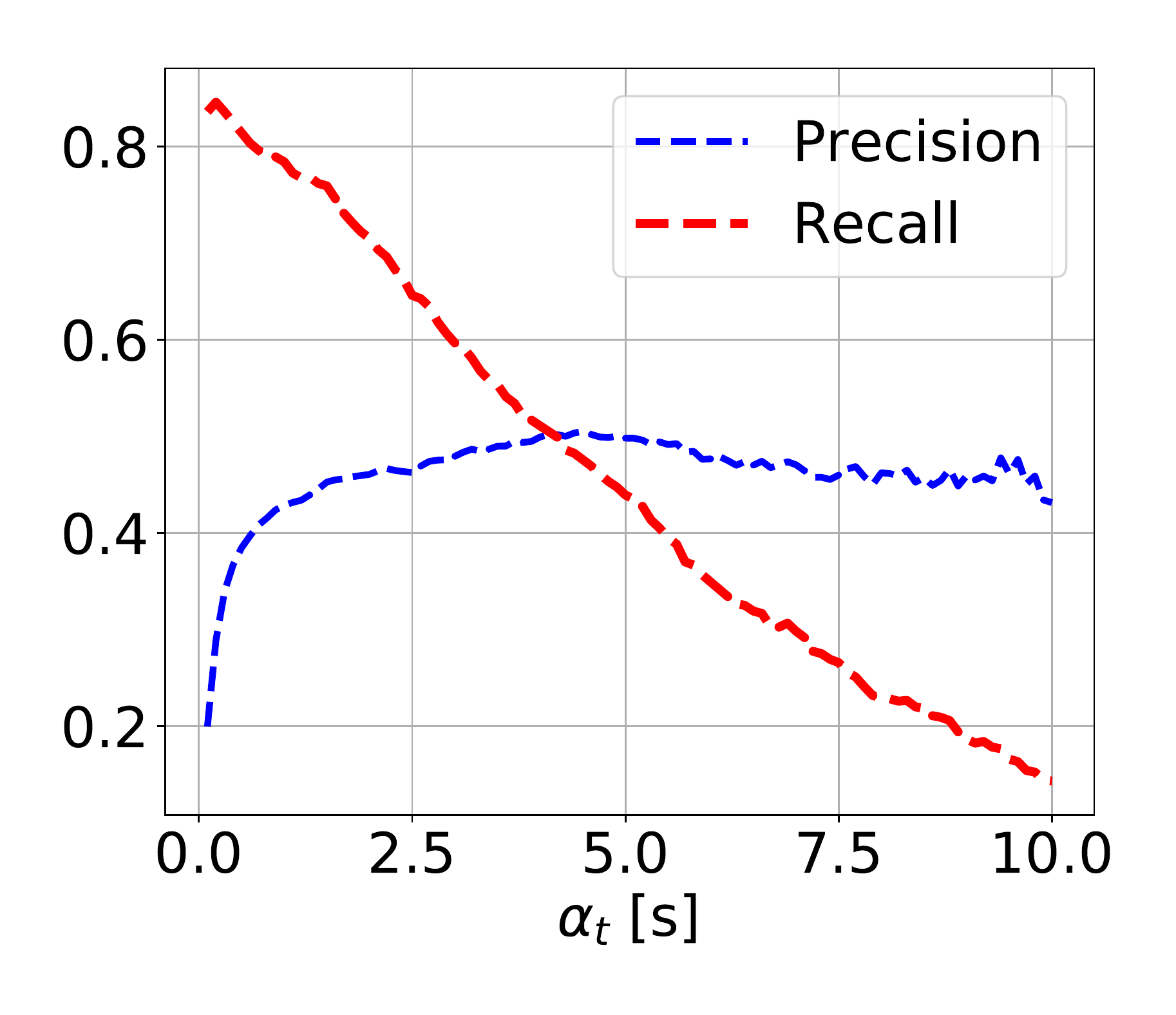}
\caption{Sensitivity analysis of the naive policy.}
\label{fig:naive}
\end{figure}

\begin{figure}
\centering
\includegraphics[width=\columnwidth]{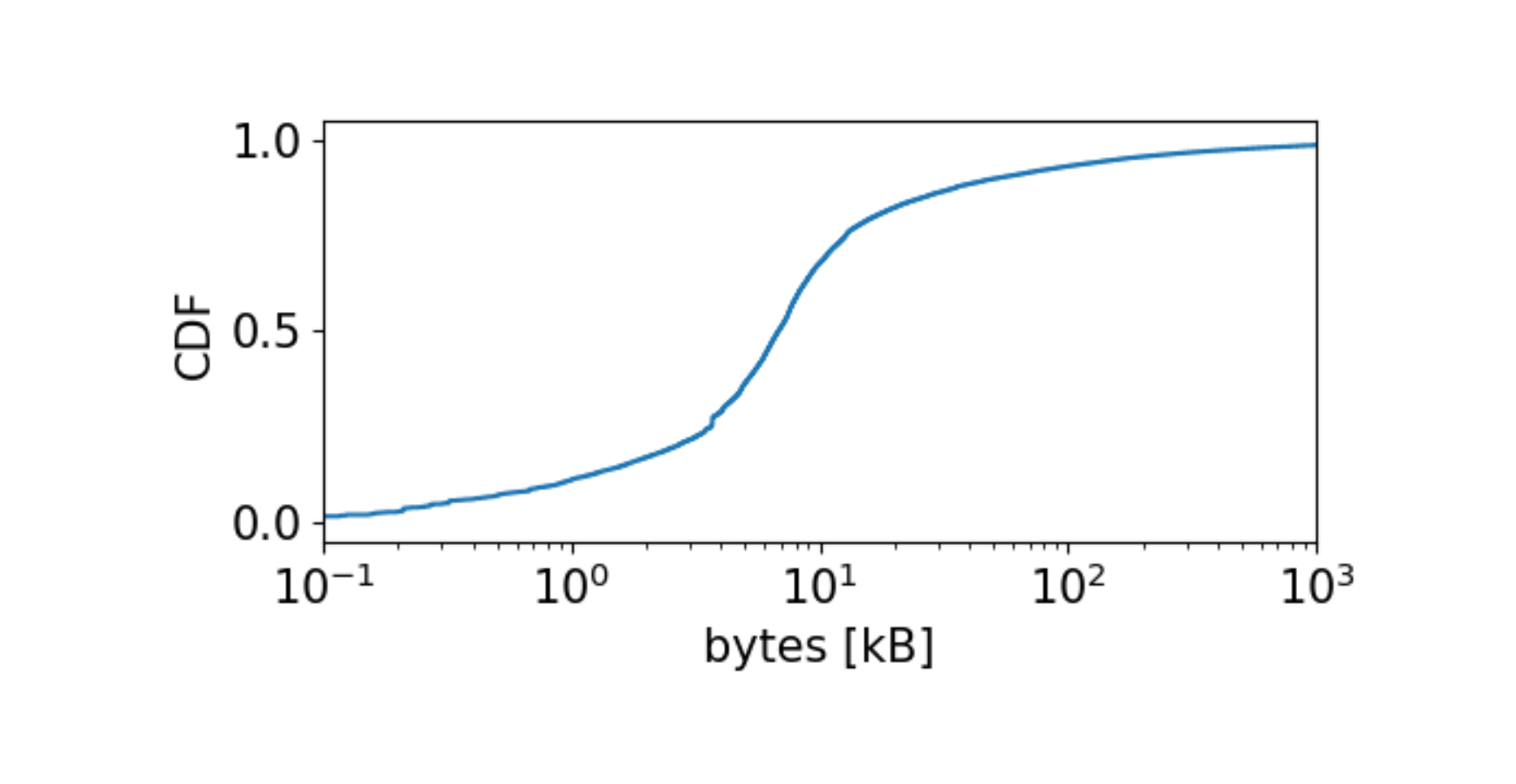}
\caption{Flow volume as seen by a large European MNO.}
\label{fig:ipfr}
\end{figure}

\begin{figure}[!t]
%\end{minipage}%
%\hspace{3pt}
%\begin{minipage}{.53\textwidth}
  \centering
  \includegraphics[width=\columnwidth]{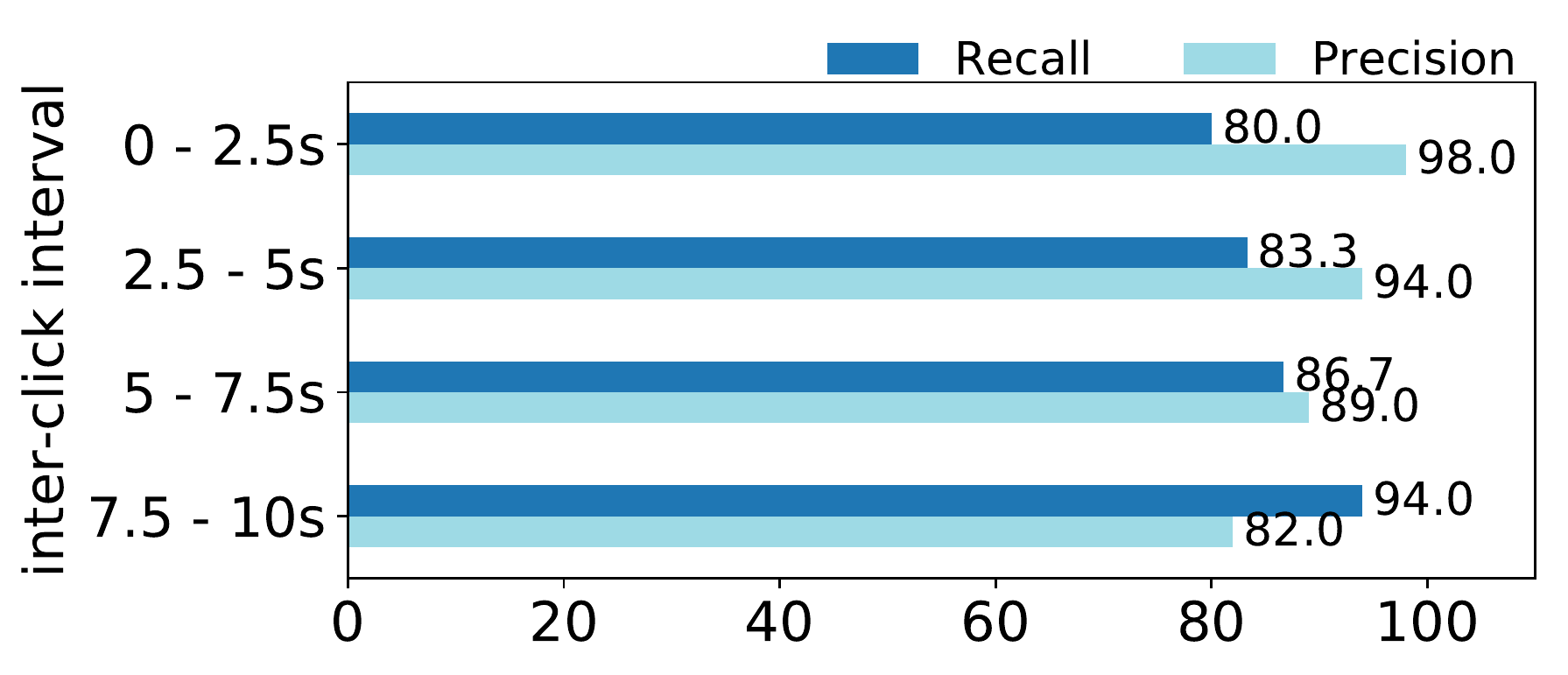}\label{fig:optimize}
%  \vspace{-10pt}
  \caption{Gradient policy sensitivity with respect to click frequency.\label{fig:clicksfrequency}}%\label{fig:interval-sensitivity}
%\end{minipage}
\end{figure}

\subsection{Validation and sensitivity analysis}

%Previous literature on mobile traffic burstness~\cite{Stober:2013:YSY,Falaki:2010:FLT}.  These works show incoming and outgoing traffic of a smartphone alternates idle periods, where almost no data is exchanged, with short peaks of high data transfer. Peaks are composed by packets that are semantically connected, \eg same TCP connection or application, and 95\% of packets arrives at most 4.5~s after their predecessor.
Our dataset contains detailed logs of the users click times, each click corresponding to the beginning of a new activity window. As such, for a given combination of thresholds, we can quantify the accuracy of the partitioning by measuring the \emph{Precision} as the fraction of partitions detected by our policies actually matching a click, and the \emph{Recall} as the fraction of clicks that are identified as activity windows by our policies. For instance, in Fig.~\ref{fig:waterfall-splitting} Precision = 1.0 and Recall = 0.25.

\noindent\textbf{Best policy.} We find the na\"ive policy being ineffective. Fig.~\ref{fig:naive} report Precision and Recall for different values of \deltat. A small threshold (\deltat $<$1s) leads to over-splitting (high Recall, but low Precision), while for larger values Recall and Precision do not go above 50\%. 
%\dpnote{changed the definition of precision and recall. Please check if it correct.}\afnote{it seems of to me, but GT should double check}
%Conversely, taking into account the presence of a burst after an idle period reduces oversplitting. Using \deltab = 5kB and \deltat = 1s both Recall and Precision are above 70\%. We selected \deltab to be the median size of a single transaction as observed in logs from a large European mobile operator (see~\cite{TECHREPORT} for details) while \deltat = 1s is considered as a minimum response time of a human response when engaging with the phone. 
Compared to na\"ive, the gradient policy, which considers bursts registered after idle periods, significantly reduces the over-splitting. By selecting \deltab = 5kB and \deltat = 1s, both Recall and Precision are kept above 70\%. We choose \deltab to be the median size of a individual transaction as observed in logs from a large European mobile operator. Results are reported in Fig.~\ref{fig:ipfr}, and are consistent with our datasets. Instead, \deltat = 1s is considered as a minimum response time of a user engaging with mobile apps.

\noindent\textbf{Interactivity.}
We also investigate whether our policy is sensible to clicks frequency.
%For this analysis we use the optimized per-category optimized values of $\Delta_b$ and $\Delta_t$ previously extracted and vary the user click interarrivals. 
%We observe interarrival times have an almost negligible \gtnote{reduced,small?} impact on the effectiveness of our algorithm. 
Fig.~\ref{fig:clicksfrequency} reports Precision and Recall across all apps when varying the frequency of clicks.
Both metrics are above 80\% in all scenarions, but there are two evident trends: Precision decreases when clicks are more sparse, while opposite is true for Recall.
%We observe that for all inter-click time ranges our algorithm correctly detects activity intervals in at least 80\% of cases. Further, we notice higher Recall and lower values of Precision as the time between two user click increases.  
%Intuitively, as the inter-click time increases, the number of activity intervals decreases, decreasing the chances of not being detected by the algorithm (\ie less false negatives, higher Recall). Conversely, as the time between two user inputs increases, there are higher chances that some fluctuations of bytes exchanged triggers out algorithm to generate a new partition, thus reducing Precision. 
%\afnote{the explanation I think is not correct, as it should not matter the number of experiments done. I think it's probably due to a sensitivity wrt the frequency, which we didn't dissect well. If we dont' find a good explanation, I would opt to cut the second set of histogram: it doesn't favor us, but we don't know exactly what is the degree of interactivity of users, so it's like discussing potentially artificial.}

\noindent\textbf{Further improvements.}
Performing a grid search to find thresholds better than the ones selected based on our intuition did not help. However, we found most of the misclassification are due to chat apps. Intuitively, as those apps typically exchange small messages (unless they are video/audio messages, or images), \deltab = 5kB is too large. Indeed, applying \deltab = 0.25kB only for this app category leads to Recall = 85\% and Precision = 88\% across all apps. Although these fine-grained optimizations could be done on a per-app basis, we argue this is unnecessary, and would be also challenging considering the large numbers of apps currently available. In fact, even if our analysis is not exhaustive, two pairs of thresholds cover a very diversified set of apps. In order to select which pairs of thresholds to use, we found that basic traffic classification techniques, based on port numbers, IP addresses, or domain names, are sufficient. For instance, chat applications use very few (and specific) domains and/or ports (\cref{sec:cpa}). 
%\dpnote{I do not understand the last two sentences: need more info or cut.}\afnote{Saying that you need multiple sets of thresholds doesn't help by itsefl, because the problem becomes: how do you know which set to use to do the split? We hint to the use of basic traffic classification (port numbers, IPs, domain name, etc.) which say which is the app (hence the class), so you know which thresholds to use.}
%\begin{figure}[!t]
%%\captionsetup{skip=-1pt}
%\centering
%%\subfigure[All apps]{
%\begin{subfigure}{0.4\linewidth}
%\includegraphics[width=1.01\columnwidth]{fig/cdf_alexa.pdf}%\label{fig:metrics-browsing-delta-cdf}
%%\subcaption{{\footnotesize Web browsing.}}
%\end{subfigure}
%%\begin{subfigure}{0.4\linewidth}
%\includegraphics[width=1.01\columnwidth]{fig/cdf_apps.pdf}%\label{fig:metrics-delta-cdf}
%%\subcaption{{\footnotesize Mobile apps.}}
%\end{subfigure}
%%\caption{{\small TDT and TDI accuracy evaluation}
%\vspace{-3mm}
%\label{fig:metrics-delta-cdf}}
%%\afnote{click $\rightarrow$ app-click; startup $\rightarrow$ app-startup}\newline
%%\afnote{replot making them square, please?}\newline
%%\afnote{add browsing / apps in plots title (rather than caption)?}
%%\afnote{improve readability}\gtnote{ok?}
%\end{figure}

\begin{figure}
\includegraphics[width=\columnwidth]{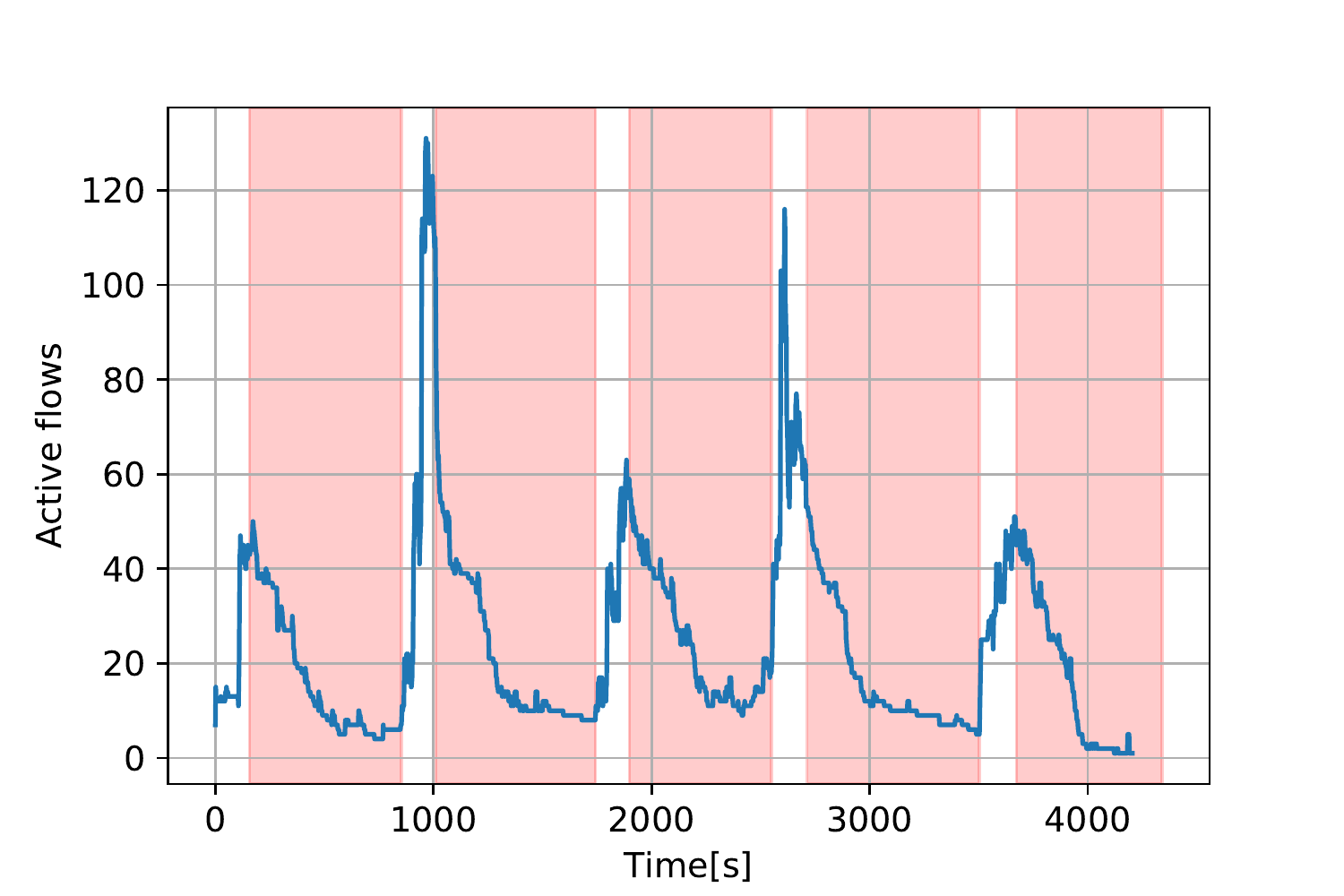}
\caption{Example of evolution of the number of active flows. Regions covered with a shaded rectangle correspond to periods when the phone screen was off.}
\label{fig:screenoff}
\end{figure}

\begin{figure}
\includegraphics[width=0.5\columnwidth]{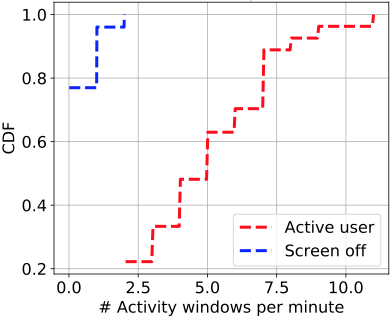}
\hspace{-5pt}
\includegraphics[width=0.5\columnwidth]{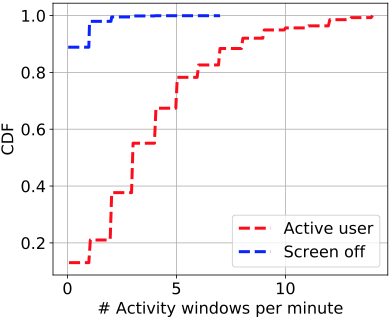}
\caption{CDF of the ratio of active windows generated per minute: synthetic user patterns (left), real users (right)
\label{fig:windowsratio}}
\end{figure}

\noindent\textbf{Background traffic.} One last aspect to consider is the impact of ``background'' traffic (notifications, emails fetch, etc.) 
%unrelated to direct user engagement, 
 on the windows partitioning accuracy.
We collected several 1-day long traces, mixing periods of activity with silence. %, and applied the na\"ive and gradient policies to detect activity windows. 
Fig.~\ref{fig:screenoff} details one of those examples, showing the number of active flows highlighting periods where the screen was on (white background) and off (shaded background). Notice how when the screen is on, \ie the user is engaging with the phone, the number of flows increases, while when the screen is off flows are progressively closed. Given the tendency to use persistent connections, flows are closed at a lower pace with respect to when they are opened.
We observe that, while the gradient policy is still sensible to background traffic, those intervals (\textit{i.e.}, with no user interaction) can be filtered out by looking at the pace at which activity windows are generated. Intuitively, when the user is active, multiple partitions are expected to be generated in a short time, while this effect is significantly reduced when only background traffic is present. Fig.~\ref{fig:windowsratio} shows this effect for both artificial clicks (monkey - left plot) and a real user (right plot). Notice how the distribution of rate at which the windows are created is macroscopically separated between periods when the screen is on and off.

\noindent\textbf{Summary.} Our results support the idea of identifying activity windows via passive measurements.  We stress that the gradient policy is a heuristic, so not meant to be perfect. Its function is to enable us to focus on traffic dynamics and CPA knowing that the portion of the traffic under analysis is likely related to user engagement, hence meaningful to be dissected. %, when it is not possible to extract fine-grained information from instrumented devices, \eg for large scale measurements.

\begin{figure*}[!t]
\centering
\includegraphics[width=0.9\columnwidth]{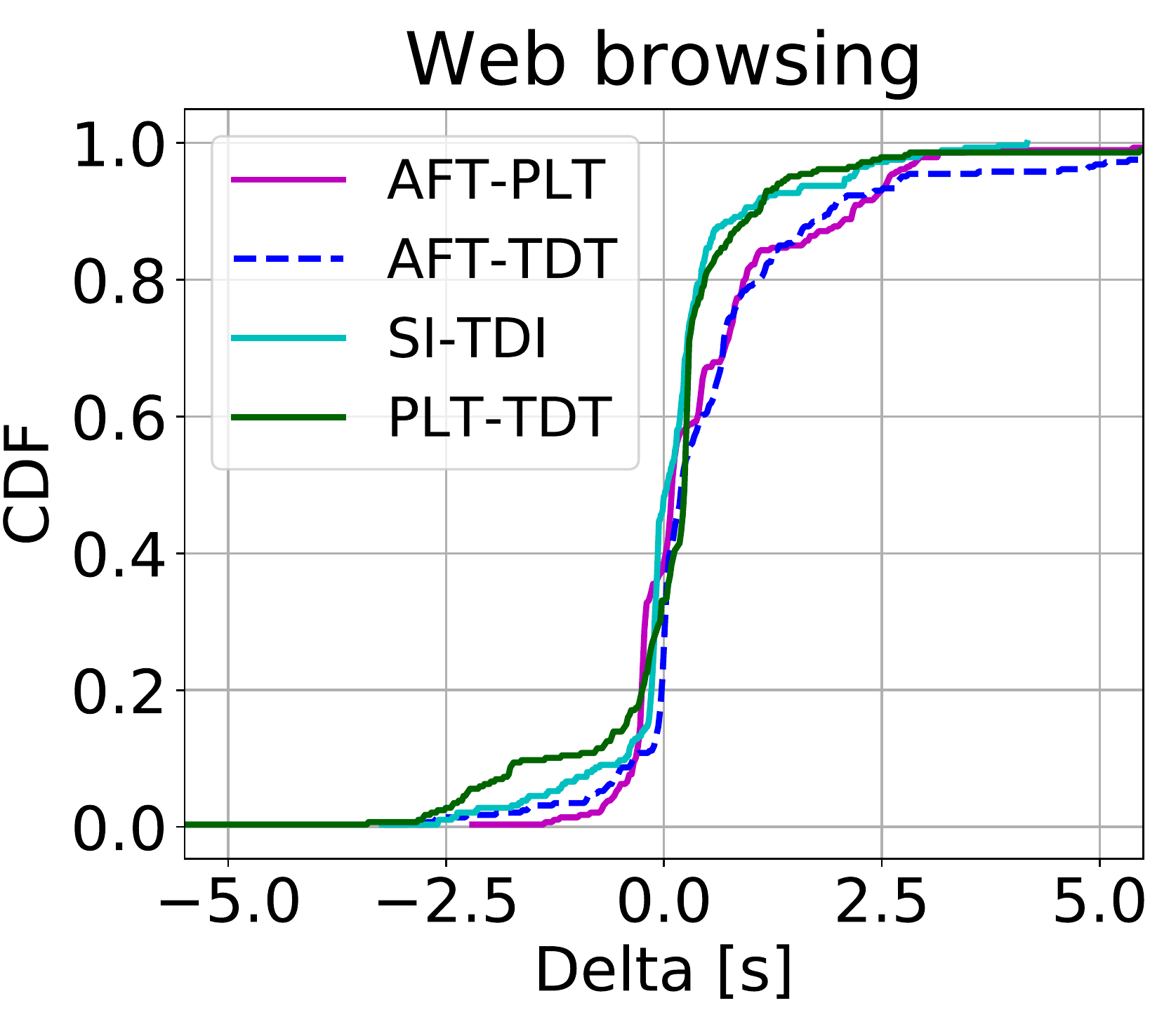}
\includegraphics[width=0.9\columnwidth]{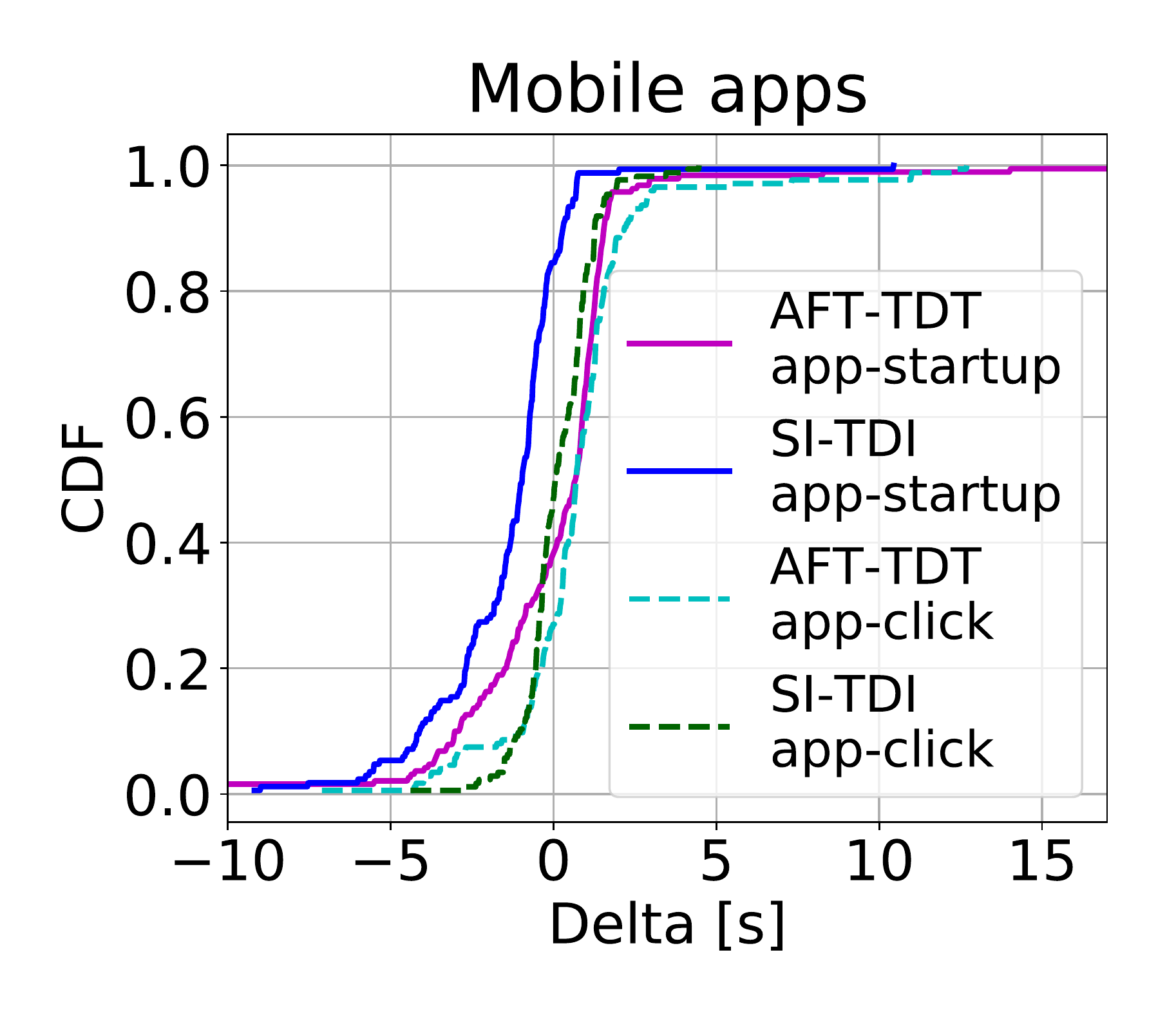}
\caption{TDT and TDI accuracy evaluation.
\label{fig:metrics-delta-cdf}}
\end{figure*}

\begin{figure*}[!t]
\includegraphics[width=1.9\columnwidth]{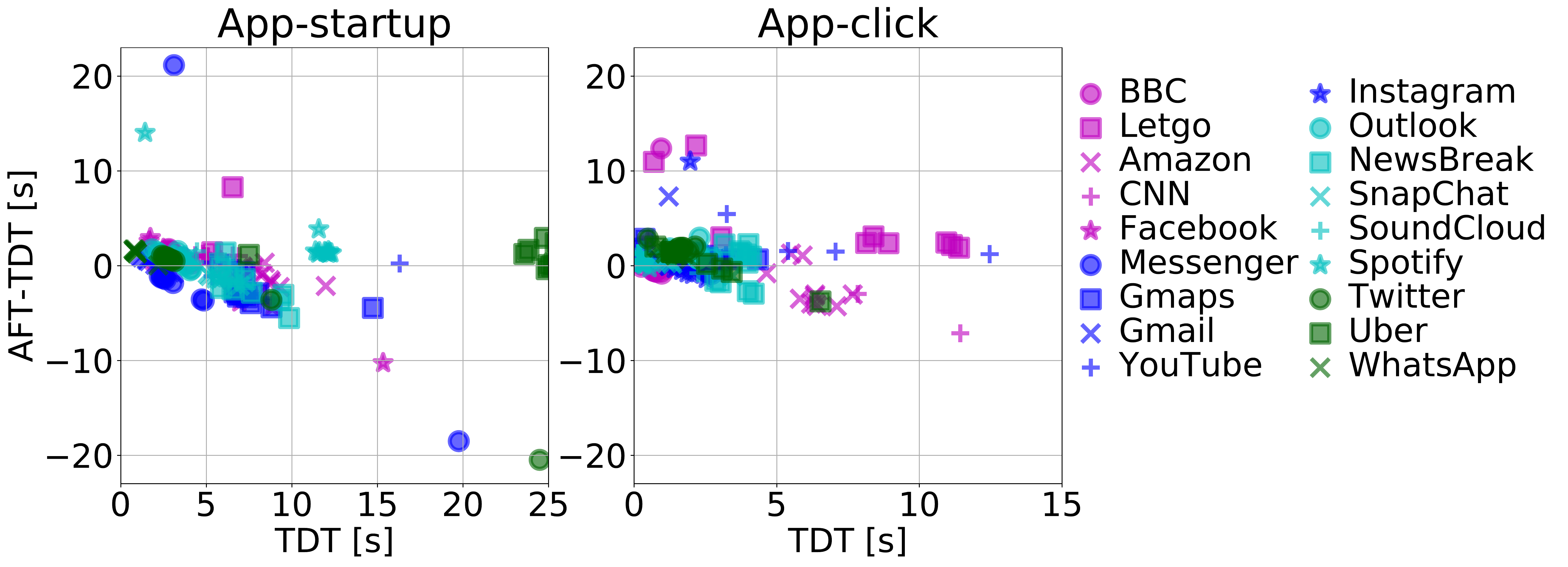}\label{fig:metrics-delta-scatter-startup}
%\hspace{-5pt}
%\includegraphics[width=0.5\columnwidth]{./fig/delta_scatter_after_startup_tma.pdf}\label{fig:metrics-delta-scatter-after-startup}
\caption{Per-app instant metrics comparison
\label{fig:scatter}}
%\afnote{can we have colors per group of apps?}
%\dpnote{Yes, but not urgent.}
\end{figure*}

\begin{figure*}[!t]
\centering
\includegraphics[width=0.33\textwidth]{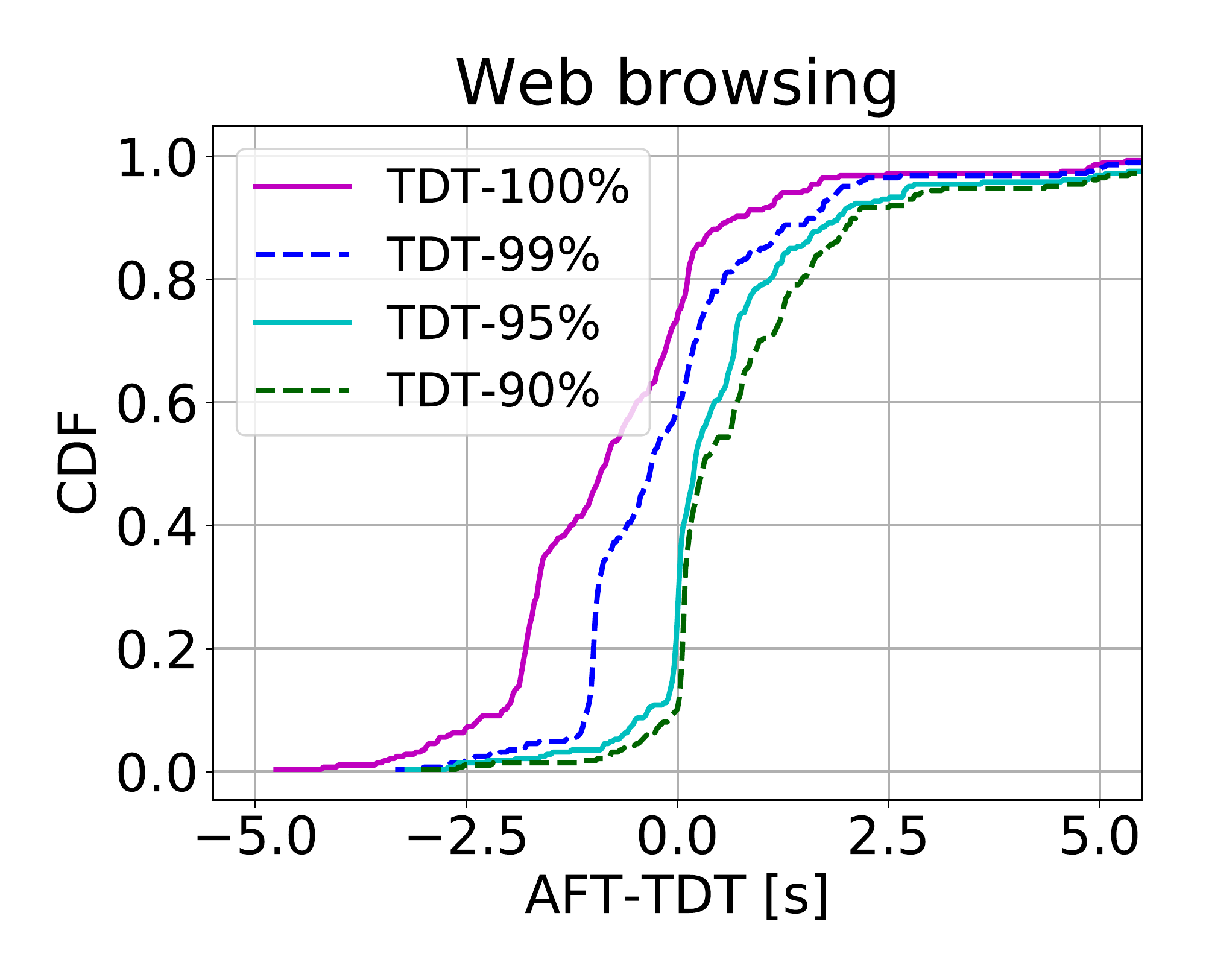}
\includegraphics[width=0.33\textwidth]{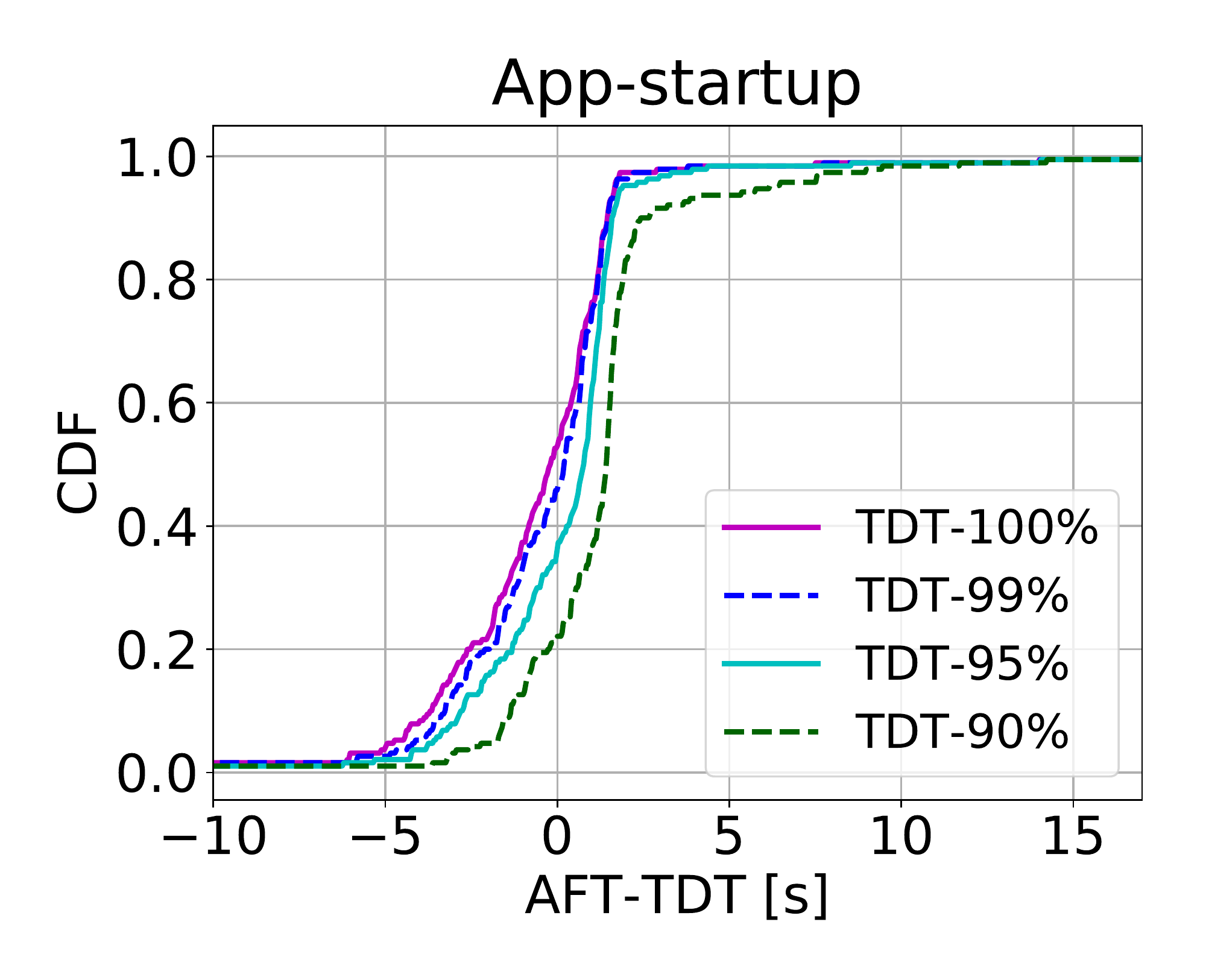}
\includegraphics[width=0.33\textwidth]{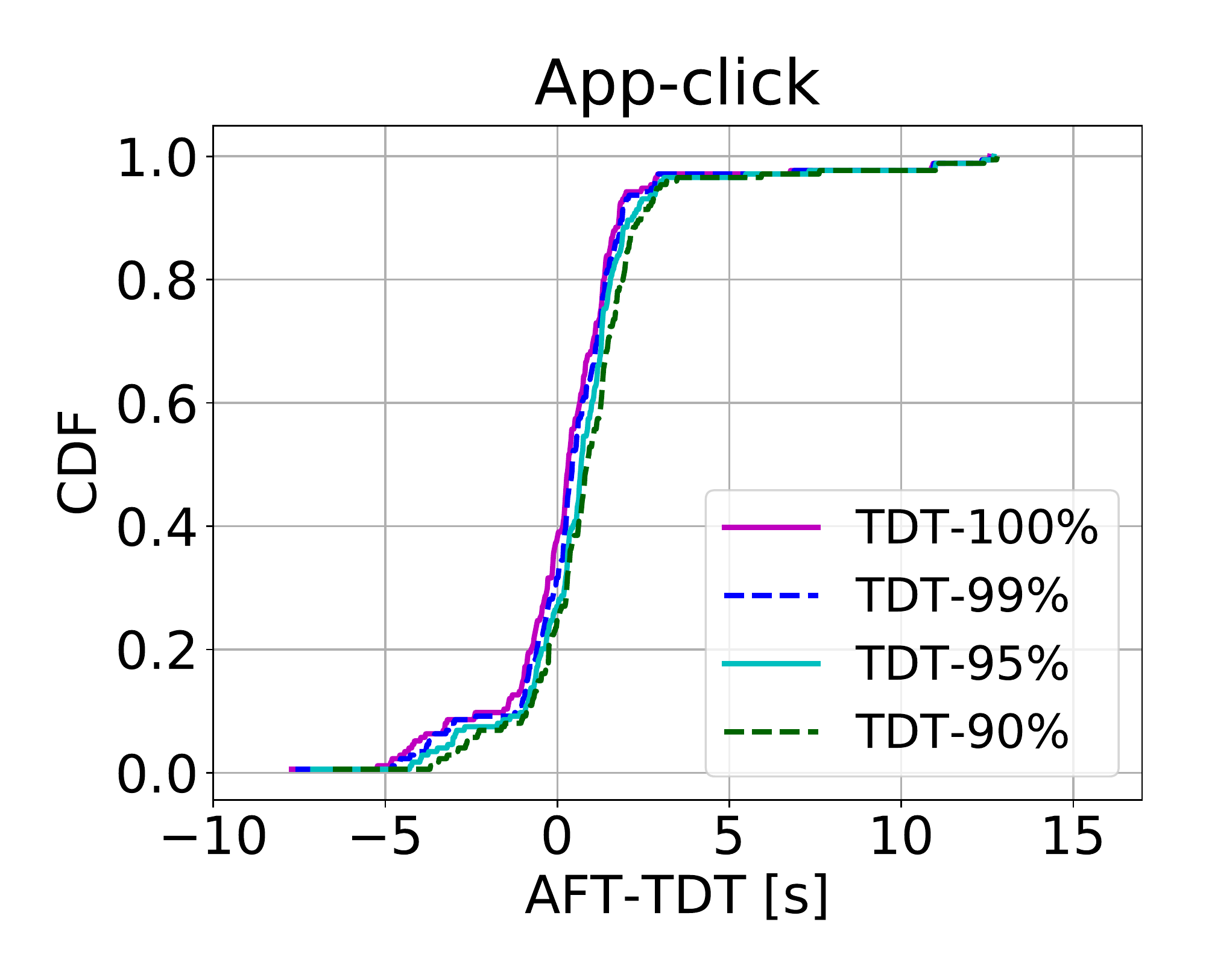}
\caption{TDT using different percentiles of the bytes cumulative\label{fig:percentiles}}
%\afnote{TODO: add final results} \gtnote{we have now 40/50 website}
\end{figure*}

\begin{figure*}[!t]
\centering
\includegraphics[width=0.51\textwidth]{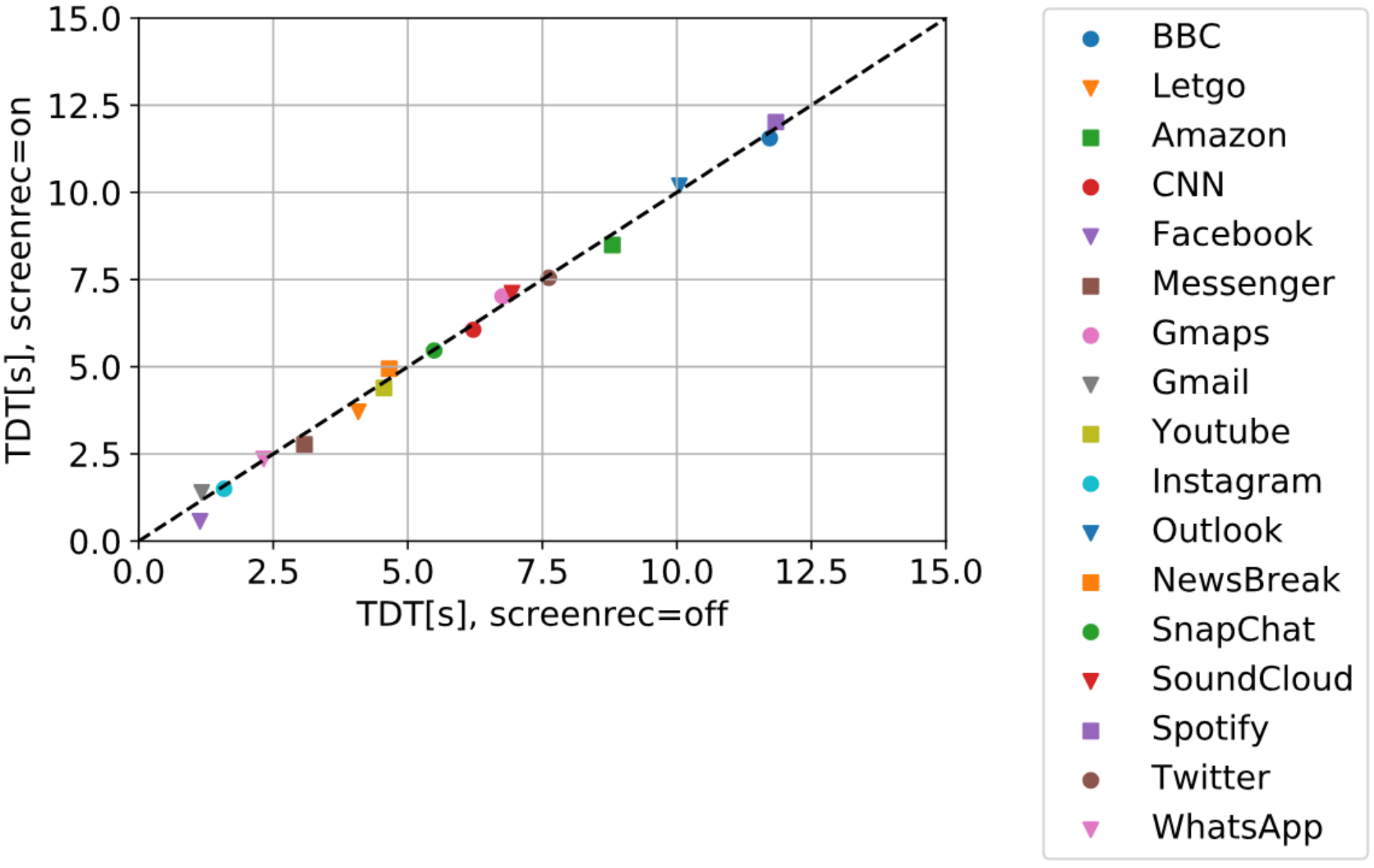}
\includegraphics[width=0.47\textwidth]{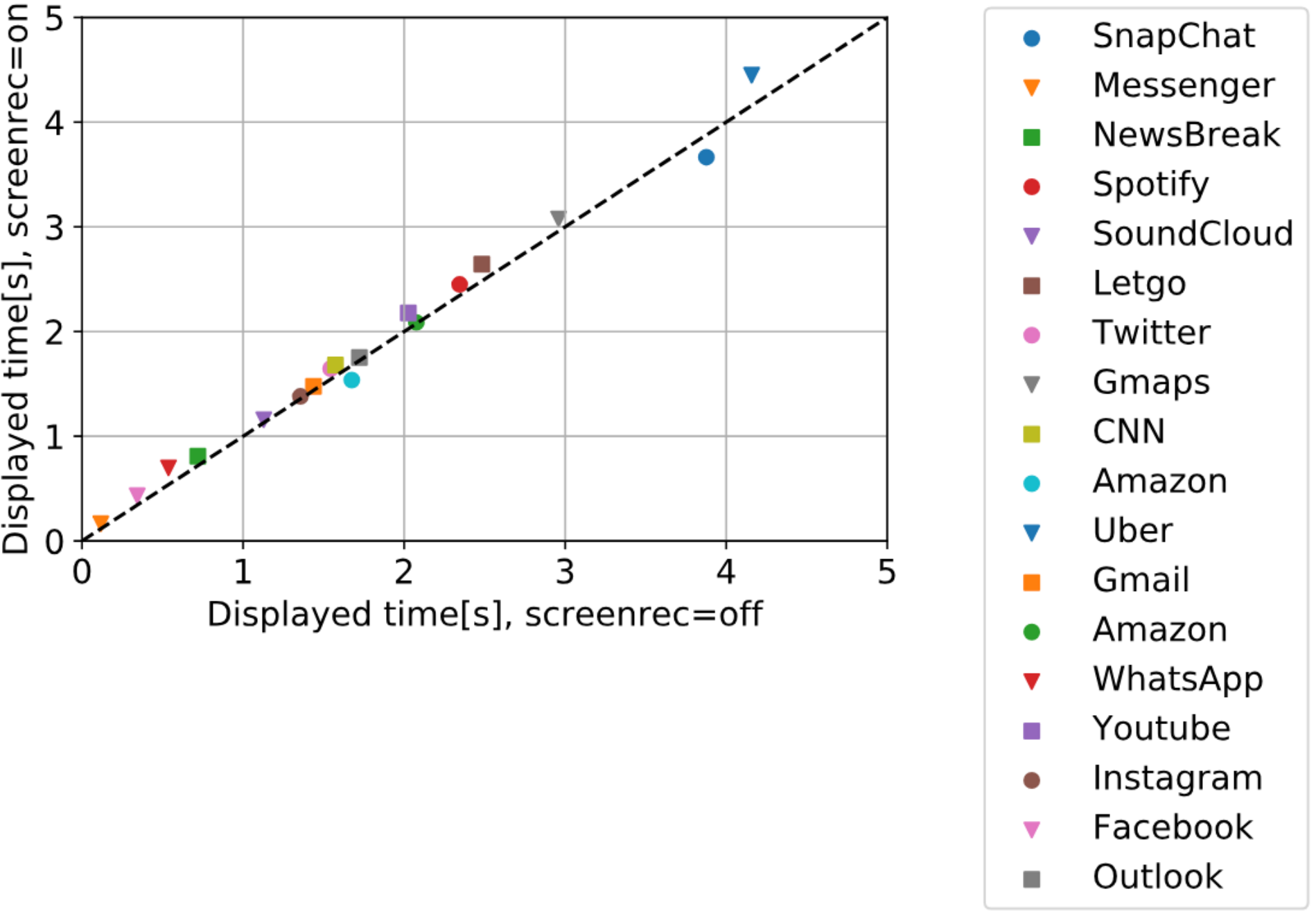}
\caption{Impact of screen recording. \label{fig:screenrec}}
%\afnote{TODO: add final results} \gtnote{we have now 40/50 website}
\end{figure*}

\section{Network Waterfall and Metrics}
\label{sec:waterfall}

%\begin{figure}[t]
%\centering
%\includegraphics[width=\columnwidth]{fig/waterfall_new2-crop}
%\caption{Example of waterfall for Letgo.}
%\label{fig:waterfall-example}
%\end{figure}

For each identified activity window, \tool creates a download waterfall detailing traffic dynamics and performance.
%reconstructs the set of network flows present in the input \texttt{pcap} file, and 

\noindent \textbf{Network waterfall.}
\tool extracts transport (L4) and application (L7) per-flow metrics. At L4, it computes aggregated statistics (\eg total duration, bytes, RTT), as well as protocol specific information (\eg TCP, QUIC, FB0 handshake duration, IP addresses, ports). At L7, \tool reports on HTTP transactions (\eg metadata from request and response headers), TLS handshake (\eg duration, if the handshake is full or fast, SNI, ALP protocols), DNS (\eg domain name, CNAMEs, query resoution time). Moreover, each flow is split into \emph{bursts} by grouping packets when interleaved by more than 2 RTTs. 
All the metrics are then represented as a download waterfall, a relevant visual aid to CPA (\cref{sec:cpa}).

\noindent \textbf{Performance metrics.}
%With the detailed measurements in a waterfall we can now define the delivery deadline for the associated activity window. 
As discussed in \cref{sec:related}, we consider AFT and SI suitable to study mobile apps traffic. However, we consider them only as baseline as we aim to avoid on-device screen recording. We are instead interested in %when these metrics cannot be measured, \eg 
%ease large scale measurements, we 
studying the reliability of objective metrics based on passive traffic measurements. %As such, we introduce two metrics: TDT and TDI.
We define the \emph{instant} metric \emph{Transport Delivery Time} (TDT) as the time between the beginning of an activity window and the 95th percentile of the whole volume exchanged in the window. We experimented with other percentiles too (see paragraph below), but the 95th resulted the more robust to long tail effect (\eg keep alive).
We also define the equivalent \emph{integral} metric \emph{Transport Delivery Index} (TDI) as $\int_0^{TDT} 1-x_B(t) dt$, where $x_B(t)$ is the percentage of total volume exchanged in the window up to time $t$. We highlight that TDI is similar to the \emph{Object Index} introduced in~\cite{bocchi2016} using TDT instead of PLT (recall that PLT does not apply for generic mobile apps \cref{sec:related}). %In the remainder of the section we compare TDT against AFT (and PLT for browsing), and TDI against SI. 
In the remainder of the section we investigate the penalties TDT and TDI introduce against the respective baselines AFT and SI. We consider also PLT as reference for browsing performance.%(and PLT whis to understand the suitability of TDT and TDI as delivery deadlines, and the penalty they introduce with respect to the reference AFT and SI.

\subsection{Evaluation}
%In order to valide our metrics, we first quantify TDT and TDI for web browsing traffic, and compare them to three state of the art metrics: \emph{Page Load Time} (PLT), \emph{Above the Fold Time} (AFT) and \emph{SpeedIndex} (SI) (cf. Sec.~\ref{sec:related}).  Then we investigate the effectiveness of our metrics in quantifying the performance of mobile applications.
%AFT capture the time it takes to render the visible portion of the screen, while SI is an integral metric of the rendering speed.
%This section validates the suitability of our performance metrics to capture the quality of the delivery performance. To this goal we compare them to QoE metrics defined for web browsing based on visual information (cf. Fig~\ref{fig:metrics-validation}), 

%\noindent \textbf{Pearson correlation.}
%Fig.~\ref{fig:metrics-correlation} reports the Pearson correlation among the different metrics when using \emph{app-startup} and \emph{app-click} datasets.
%Notice how all metric pairs present a strong correlation:
%TDT shows a 0.84 correlation with the related state-of-the art instant metric AFT; the correlation remains significant (0.71) when considering integral metrics too (\ie TDI and SI. %. Finally, TDT is highly correlated with TDI and AFT is well correlated with SI. 
%Although the plot reports aggregated results, there are not significant differences among scenarios and applications.

\noindent \textbf{Web Browsing.}
Fig.~\ref{fig:metrics-delta-cdf}(left) reports the Cumulative Distribution Function (CDF) of the deltas AFT-TDT and SI-TDI for \topalexa dataset. Both are well centered around zero, but TDI is a better proxy of SI than TDT is for AFT. Notice however that AFT-PLT presents a similar distribution as AFT-TDT. In other words, if PLT is the most popular metric to measure web performance, TDT is at least comparable. This is further corroborated considering PLT-TDT which presents a distribution well centered around zero.
%However, AFT-PLT and AFT-TDT distributions are slightly negative. Indeed, when the visible part of a webpage is completely rendered some bytes are still downloaded in background~\cite{aft,eyeorg}, hence TDI and PLT are larger than AFT. \afnote{This is why AFT is considered the best metric to reflect rendering effect, despite PLT is the metric commonly adopted despite its known issues. Clearly TDT has differences from AFT, but at least as good as PLT --- can we prove something like this?} 
%\afnote{why we don't have PLT-TDT? Don't remember if the did it or not...but could support the idea that TDT is not worst than PLT (which is what people use in the end), even if PLT is notoriously worst than AFT.}
%Differently, SI-TDI distribution is sharply centred around zero. In other words, the largest amount of bytes are retrieved when the page rendering is in progress, hence TDI is close to SI.
%\afnote{I feel like we are underselling the fact that the integral metrics are better than punctual ones...}

\noindent \textbf{Aggregate apps traffic.}
Fig.~\ref{fig:metrics-delta-cdf}(right) reports the CDFs of AFT-TDT and SI-TDI deltas for both \appstartup and \appclick datasets. 
All curves are well centered around zero, but \appstartup CDFs present a heavier negative tail. %In this scenario, the delta between SI and TDI is negative, while the delta between the corresponding time instant metrics, \ie $TDI - AFT$, is around zero. 
%This is due to the fact that the bulk of images and information visualized on the screen is either already available or quickly downloaded. This does not happen on users clicks as the visualized information changes when data is retrieved only, \ie delta between both time instant and time integral metrics is centered around zero. 
This resembles what was observed for browsing, \ie at startup more content is downloaded than what is required for the visualization, so TDT and TDI can over-estimate rendering deadlines. TDI is more sensible to this effect, while for 75\% of the experiments TDT generates a $\pm$1.3s error at most.
%We stress that this bias is also related to the nature of the \appstartup experiments, as we flush apps cache between different runs to enforce a cold start scenario. On average, this leads to dowload 25\% more volume in \appstartup than for \appclick. This is an extreme scenario, as in practice this condition should not happen often.
%\afnote{can we say that even when in app-click we are still downloading some traffic? Probably the number in Table1 can help, but if I'm not wrong those are the bytes over the critical path, right? We need to say here that the avg different between startup and click is X\% ... with hopefully a small number?} \gtnote{On average, 25\% bytes less are downloaded for app-click compared to app-startup}

%\noindent \textbf{Mobile apps breakdown.}
%Figure~\ref{fig:metrics-delta-cdf} also shows 20\% of experiments for which the delta is larger than 5s. 
\noindent\textbf{Per-app traffic.} To further investigate the deviations between instant metrics, Fig.~\ref{fig:scatter} 
reports the deltas AFT-TDT as a function of TDT for each individual app. %Each point is a different run, and we use different point types to distinguish between apps. 
Considering \appstartup (left plot), besides a few outliers, all apps present similar behavior, with variable deadlines in absolute scale, but TDT is triggered slightly after AFT as already observed in Fig.\ref{fig:metrics-delta-cdf}(right). %. peronly Uber \gtnote{this needs to change, since for Uber the delta is now close to 0} presents a different behavior. Indeed, the app downloads 60\% volume after the AFT, hence affecting the overall AFT-TDT distribution in Fig.~\ref{fig:metrics-delta-cdf}(left). %Further, Uber has large values of both TDT and AFT at startup. 
For \appclick (right plot) errors are further reduced, with only Amazon showing larger penalties.
%The same analysis is repeated for \emph{app-click} dataset in Fig.~\ref{fig:metrics-delta-scatter-after-startup}.
%Results are more concentrated around zero, although Amazon presents a different behavior.
%We observe a similar behavior, despite less significant, for Amazon app on users clicks. 
%We notice indeed that the Amazon app loads content outside the visible portion of the screen, which justifies the observed deltas. %e differe visible to the user is retrieved once the above the fold portion is already visible. 
%Although not shown, these considerations hold for the delta among time integral metrics too.

\noindent\textbf{TDT sensitivity to percentiles.} 
TDT and TDI capture the progress of the download by means of a percentile. The analysis previously reported refers to the 95th percentile of the volume transfered within an activity window, but other values are possible. To better investigate the sensitivity of selecting the percentile to use, Fig.~\ref{fig:percentiles} reports the CDFs of the delta AFT-TDT for web browsing, \appstartup, and \appclick respectively. Intuitively, selecting a high percentile can expose the passive metrics to pre-fetching, \ie the delivery deadline triggers too late due to content downloaded even if not required for the rendering. Conversely, a small value instead cause the opposite effect, \ie the deadline triggers too early. Both effects are clearly visible in Fig~\ref{fig:percentiles} CDFs. Considering browsing (left plot) waiting until all content is downloaded (TDT-100\%) results is a macroscopic delay with respect to AFT, while 95th and 90th present fairly similar performance. Considering app traffic, \appstartup (middle plot), high percetiles give similar performance, while the 90th percentile is less accurate; \appclick (right plot) presents very little differences. Overall, we selected the 95th percentile to define TDT and TDI as it provides more consistent performance across the different types of traffic.

\noindent\textbf{Impact of screen recording.}
As the video screen record can be resource demanding, it can bias the measurement of AFT and SI, as well as our defined deadline TDT and TDI.
To investigate on this, we consider the apps startup, and run 10 experiments for each app with and without \texttt{Android screenrecord} enabled. For each experiment we collect two metrics: TDT over the app startup activity window, and %which is used to evaluate whether the screen record slows the byte exchange down. The second metric is the 
the \texttt{Displayed} time\footnote{\url{https://developer.android.com/topic/performance/vitals/launch-time}} provided by \texttt{Android Activity Manager}\footnote{\url{https://developer.android.com/reference/android/app/ActivityManager}} via \texttt{logcat} for those activities involved in the app startup (\eg launching the process, initializing the objects, creating the activity, inflating the layout, and drawing the application for the first time). In other words, we are interested in understanding if (and by how much) these deadline changes in the presence of the screen record with respect to the bytes exchanged, as well as the dynamics of the apps.
In Fig~\ref{fig:screenrec} we compare the median TDT (left) and Displayed times (right) across apps and experiments. Notice how the points are well distributed over the bisect line, with only two corner cases for the Display time (Uber and Whatsapp). With an error generally lower than 1\%, we can conclude that for the apps under study the impact of the screen record is marginal.

\noindent\textbf{Summary.} The analysis shows that metrics purely based on passive traffic monitoring are a reasonable approximation of AFT and SI, and at least as good as popular metrics such as PLT. This brings visibility on apps dynamics when AFT and SI cannot be measured, and more broadly they can significantly simplify QoE/performance analysis. There are clearly some corner cases and occasional outliers, as not all apps behave the same, but our analysis shows that TDT and TDI are reasonable heuristics to qualitatively capture delivery deadlines.

%\begin{figure}[t]
%\includegraphics{fig/correlation_graph.pdf}
%\caption{Pearson correlation among the different performance metrics. All scenarios and applications.}\label{fig:metrics-correlation}
%\end{figure}

%We now analyze the waterfall construction procedure. HTTP transactions, DNS requests and responses, and TCP/TLS/QUIC handshakes can be exactly identified as all information required are available in \texttt{pcap} files.
%However, we still need to analyze the identification procedure for TLS/QUIC transactions. Differently from the previous section where we have a ground truth to evaluate the effectiveness of our algorithm, information on actual application layer transactions is not available. We therefore study the impact of the parameter $T_S$ on the number of transactions and transaction size for a set of application at startup in Fig.~\ref{fig:tls-segmentation}. For readability, we report results for one application per category in the startup scenario; other applications the application usage scenario follow the same trend. 

%For all applications, we observe a sharp decrease in the number of transaction and transaction size as $T_S$ increases, followed by a plateau. \dpnote{Finish comment but I have an issue. do not know exactly what to say as I do not clearly see which value to pick}
%\dpnote{Modify comments according to figure} 

\begin{figure}[!t]
\rule{55pt}{0pt}\includegraphics[width=0.65\columnwidth]{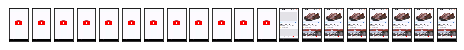}\\
\includegraphics[width=\columnwidth]{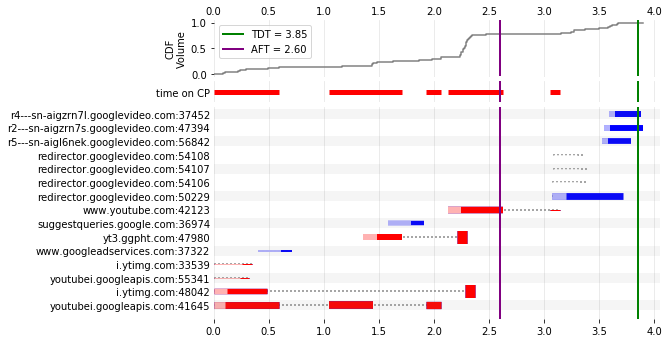}\\
\rule{19pt}{0pt}\includegraphics[width=0.9\columnwidth]{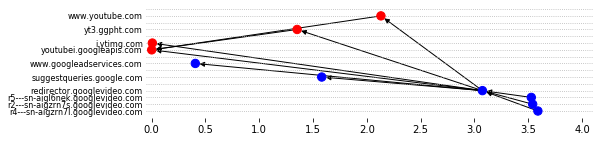}
\caption{YouTube startup waterfall.\label{fig:youtube}} 
\end{figure}

\begin{figure}[!t]
\includegraphics[width=\columnwidth]{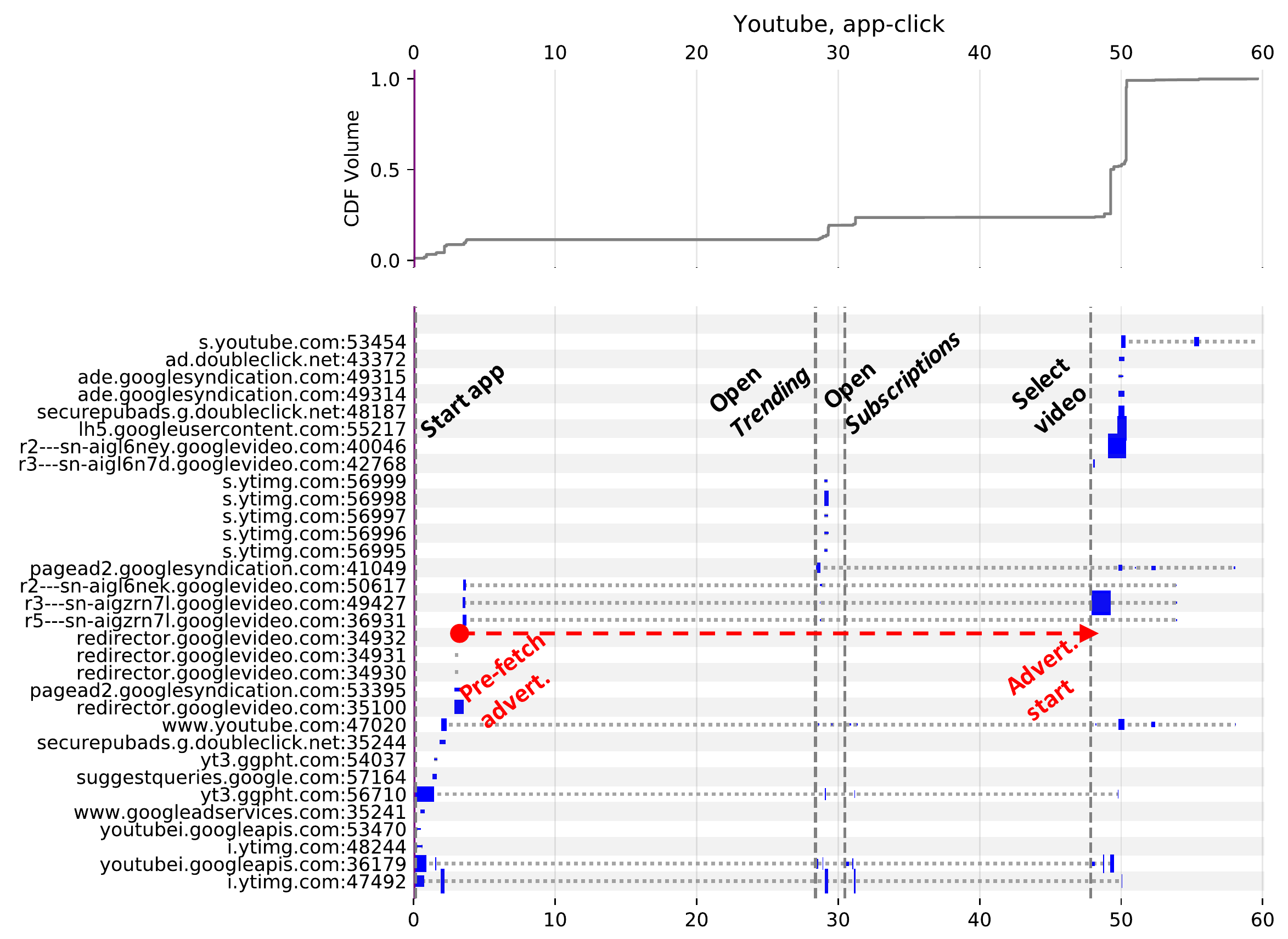}
\caption{Youtube \appclick traffic flows. The red arrow denotes possible prefetching of the advertisement video. \label{fig:youtube-multi}}
\end{figure}

\begin{figure}[!t]
\rule{55pt}{0pt}\includegraphics[width=0.65\columnwidth]{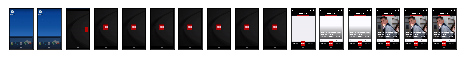}\\
\includegraphics[width=\columnwidth]{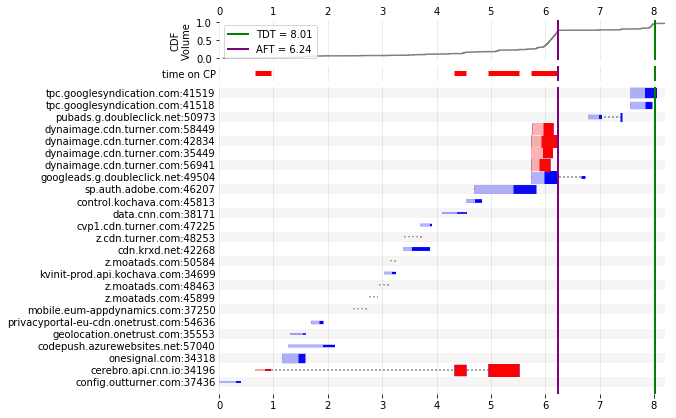}
\rule{28pt}{0pt}\includegraphics[width=0.89\columnwidth]{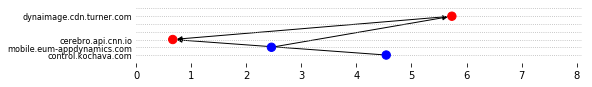}
\caption{CNN startup waterfall.\label{fig:cnn}} 
\end{figure}

\begin{figure}[!t]
\rule{10pt}{0pt}\includegraphics[width=0.65\columnwidth]{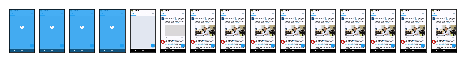}\\
\includegraphics[width=\columnwidth]{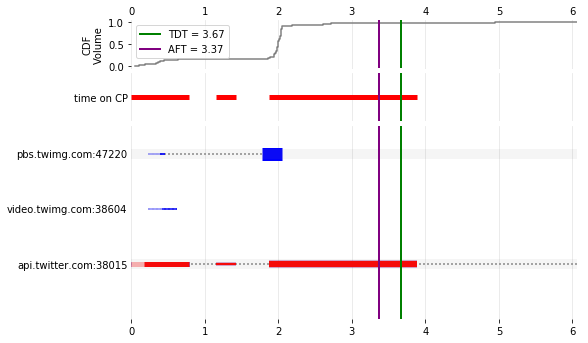}
\rule{1pt}{0pt}\includegraphics[width=0.9\columnwidth]{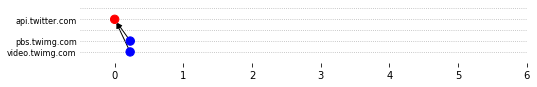}
\caption{Twitter startup waterfall.\label{fig:twitter}}
\end{figure}

\section{Critical Path Analysis}
\label{sec:cpa}

%The last step of \tool is the identification of the critical path of every activity window, \ie network flows that are bottleneck in the delivery of application resources.  Our rationale is to first run every application multiple times, each time shaping the traffic associated to one domain name in the activity interval, using the \textit{tc} traffic control tool on the test device.  Then, we identify from those experiments which domains have an actual impact on the performance metric TDT.  The result is the set of {\it critical} domains, i.e. domains for which a delay in the download/data-exchange will affect the overall delivery performance. Flows belonging to such domains constitute the critical path for the activity window under analysis.

CPA tools for browsing define the critical path based on a dependency graph capturing the relations between objects downloaded (\cref{sec:related-critical}). This graph is constructed ``passively'' exploiting the DOM built by the browser when rendering the webpage; however this technique is not applicable to generic mobile apps.
Therefore, to discover critical traffic, \tool uses an ``active'' approach based on traffic throttling.
We use the \texttt{tc} utility to throttle one domain at a time to 1kb/s, and test the impact on the activity window delivery deadline. 
In particular, for each throttling scenario we perform 10 runs applying a p-value test (with $0.05$ as significance level) to accept or reject the null hypothesis: a domain is critical if the deadline is always delayed across runs. Likewise, a similar test is applied to discover dependencies among domains (\ie by delaying domain A also domain B is delayed).

Overall, we define \emph{Critical Set} (CS) as the set of domains impacting the delivery deadline, and we use it to create a dependency graph among domains. We define \emph{Critical Path} (CP) as the whole set of flows generated by the CS. %This is a more ``relaxed'' definition than what commonly used in literature (\cref{sec:related}) as we are interested in understanding the whole traffic dynamics, rather than pinpointing specific activities, or parts of the dependency graph.
In other words, similarly to Lighthouse, \tool CP is defined based only on network traffic, but it captures the whole traffic activities of a flow, rather than pinpointing specific objects/requests.
It follows that the \emph{time on CP} is the sum of time intervals where at least 1 critical flow is active. 
%, \ie total time for which any data is transmitted or received to/from critical domains. 
%the definition is more ``relaxed'' than what is commonly adopted (\cref{sec:related}).
%Based on the set, we define the \emph{Critical Path} (CP) as the dependency graph generated by the CS. The \emph{critical traffic} is the 
%from the ``union'' of all critical traffic: if $flow_1$ has a burst between ($t_0$, $t_1$) and $flow_2$ has a burst between ($t_2$ and $t_3$), and flows are overlapped with $t_0 < t_2 < t_1 < t_3$, this create an overall critical time ($t_0, t_3$).
%as composed by the flows associated to domains in the CS, while the time spent on the critical path is the sum of intervals for which at least one of these flows is active, \ie total time for which any data is transmitted or received to/from critical domains. 
In the remainder of the section, we first present some examples of CPA on specific apps. Then, we discuss traffic properties across apps.

\subsection{Dissecting individual apps traffic}
Fig.~\ref{fig:youtube}, Fig.~\ref{fig:cnn}, and Fig.~\ref{fig:twitter} details the startup traffic dynamics for YouTube, CNN, and Twitter apps respectively by stacking 6 views of the traffic: dependency graph, download waterfall, time on CP, CDF of the bytes exchanged, and a film strip showing the screen rendering progress. The dependency graphs show only domains having at least one dependency. In the download waterfall each row corresponds to a different flow (labeled with domain and destination port). Horizontal lines show bursts carried by flows (\cref{sec:waterfall}), colored red if found critical (blue otherwise), while dotted lines indicate idle periods. Saturated colors reflect exchange of data, while pale ones correspond to DNS and handshakes (TCP, TLS, or QUIC).
Finally, two vertical lines mark the AFT and TDT deadlines.
% On top of the waterfall we l is summarized showing the time on CP, and the CDF of the overall volume. We also add a set of screenshots portraying the evolution of the screen rendering. 

\noindent\textbf{YouTube.} Focusing on YouTube, the traffic before the AFT is almost entirely critical.
This is composed of a mix of images (\emph{i.ytimg.com} handles video thumbnails, while \emph{yt3.ggpht.com} handles user related content such as avatars), control, and other structural elements of the app (\eg fonts, javascripts). %\emph{www.youtube.com}, \emph{youtubei.googleapis.com}). 
The download idle times hint to rendering cycles (fetch $\rightarrow$ process $\rightarrow$ render $\rightarrow$ iterate), %, with the final rendered content is presented to the user with a fading transition. 
as also confirmed by the film strip showing a ``dummy'' loading screen used to hide the actual rendering process.
TDT is delayed due to video pre-fetching. %: \emph{redirector.google.com} is used to load video properties~\cite{XXX}, while \emph{xyz.googlevideo.com} are video caches. 
This is confirmed by \appclick, where we observe the portion of video left being delivered on the already opened flows when the playback is triggered. An example is shown in Fig.~\ref{fig:youtube-multi}. After 4 requests to \emph{redirector.google.com}, three flows are opened towards video caches (\emph{r5---sn-aigzrn7l.googlevideo.com}, \emph{r3---sn-aigzrn7l.googlevideo.com}, and \emph{r2---sn-aigl6nek.googlevideo.com}). At around 28s and 30s, the {\it trending} videos tab and the {\it subscriptions} tab are opened, respectively. Finally, at 47s the video playback is triggered. Notice however how a ``blob'' of content is carried through the connection opened earlier and correspond to a video ad, while the actual video is downloaded from a different video cache (\emph{r2---sn-aigl6ney.googlevideo.com}).

%\begin{figure}[t]
%\centering
%\includegraphics[width=0.8\columnwidth]{fig/uber_waterfalls/uber_fr}\newline
%\includegraphics[width=0.8\columnwidth]{fig/uber_waterfalls/uber_wf}
%\caption{Example of Uber startup download waterfall.
%\label{fig:uber-startup}}
%\end{figure}

%\begin{figure}[t]
%\centering
%\includegraphics[width=0.8\columnwidth]{fig/cnn_waterfalls/cnn_fr}\newline
%\includegraphics[width=0.8\columnwidth]{fig/cnn_waterfalls/cnn_wf}
%\caption{Example of CNN startup download waterfall.
%\label{fig:cnn-startup}}
%\end{figure}

\noindent\textbf{CNN.} Differently from YouTube, the majority of the traffic for the CNN app is not critical. 
After contacting \emph{cerebro.api.cnn.io} (possibly a control domain), there are about 3s busy with only 3rd party and ads services communications, none of which is critical.
Finally the control goes back to \emph{cerebro.api.cnn.io} which triggers the rest of the critical traffic (\emph{dynaimage.cdn.turner.com}). 
As for YouTube, rendering phases are possibly hidden by the loading screen, but more interesting is the macroscopic impact of 3rd party traffic which accounts for 55\% of the overall deadline.

\noindent\textbf{Twitter.} The Twitter app instead has a very simple waterfall: only 3 flows, all twitter related, with only 1 being critical. We interpret this minimalist approach as an explicit design choice, but it would be interesting to know if applying content sharding and a few more flows could further reduce loading latency.

\begin{table*}[!t]
\centering

\caption{Critical path traffic characteristics.
\label{tab:critical}}

\begin{tabular}{
    r@{\hspace{2pt}}
    |@{\hspace{0pt}} r @{\hspace{1pt}} r @{\hspace{1pt}}
    |@{\hspace{0pt}} r @{\hspace{1pt}} r @{\hspace{1pt}}
    |@{\hspace{0pt}} r @{\hspace{1pt}} r @{\hspace{1pt}}
    |@{\hspace{0pt}} r @{\hspace{1pt}} r @{\hspace{1pt}}
    |@{\hspace{0pt}} r 
     @{\hspace{1pt}} r
     @{\hspace{1pt}} r
     @{\hspace{1pt}}|
    }
\multicolumn{1}{c}{} &
\multicolumn{11}{c}{App-startup} 
\\
\hhline{~-----------}
\multicolumn{1}{@{\hspace{0pt}}c@{\hspace{1pt}}}{} &
\multicolumn{2}{|@{\hspace{0pt}}c@{\hspace{1pt}}}{fl.} &
\multicolumn{2}{|@{\hspace{0pt}}c@{\hspace{1pt}}}{dom.} &
\multicolumn{2}{|@{\hspace{0pt}}c@{\hspace{1pt}}}{vol.[kB]} &
\multicolumn{2}{|@{\hspace{0pt}}c@{\hspace{1pt}}}{TC[s]} &
\multicolumn{3}{|@{\hspace{0pt}}c@{\hspace{1pt}}|}{TC break [\%]}
\\
%App         & all   & CP    & all  & CP              \%Dns & \%Tcp & \%Tls/QUIC & \%Data \\ \hline
%Letgo       & 736   & 715   & 3.6  & 0.65            & 6     & 6     & 25         & 63     \\ \hline
%Amazon      & 1560  & 1490  & 8.22 & 6.98            & 0.4   & 7.8   & 22.8       & 69     \\ \hline
%CNN         & 80    & 25    & 7.06 & 2.69            & 15    & 15.8  & 3.8        & 65     \\ \hline
%Gmaps       & 880   & 870   & 7.3  & 4.2              & 0     & 12    & 25         & 63     \\ \hline
%Youtube     & 151   & 127   & 6.59 & 3.03            & 5.8   & 9     & 16         & 68.7   \\ \hline
%Outlook     & 22    & 20    & 2.7  & 2.15             & 3     & 10    & 22         & 65     \\ \hline
%NewsBreak   & 165   & 152   & 7.1  & 4.5              & 0     & 8     & 12         & 80     \\ \hline
%SoundCloud  & 726   & 715   & 10.6 & 8.15            & 0     & 6     & 16         & 78     \\ \hline
%Spotify     & 82    & 78    & 7.8  & 4.6              & 22    & 9     & 6          & 62     \\ \hline
%Uber        & 245   & 61    & 8.3  & 5.36            & 0     & 8     & 20         & 72     \\ \hline
%BBC         & 102   & 98    & 2.46 & 0.52            & 0     & 29    & 0          & 71     \\ \hline
             &abs  &\%   & abs &\%   &  abs   &\%    &abs  &\%    & dns   & hshake        & data \\
\hhline{~-----------}
Twitter     &5   &38   &1  &13   &33    &79   &4   &77    &0    &32    &68\\   %&14    &18    &68\\
Facebook    &2   &40   &2  &40   &836   &97   &9   &61    &1    &6.3   &92.7\\ %&0.2   &6.1   &92.7\\
Instagram   &9   &56   &2  &25   &1108  &97   &4   &80    &0    &11.6  &88.4\\ %0.2   &11.4  &88.4\\
\hhline{~-----------}
Whatsapp    &2   &100  &1  &100  &4     &100  &1   &100   &9    &6     &85\\   %&6     &0     &85\\
Snapchat    &8   &80   &4  &50   &2802  &91   &11  &70    &3    &23    &74\\   %&4     &19    &74\\
Messenger   &4   &57   &3  &50   &86    &72   &2   &63    &0    &31.8  &68.2\\ %&12.5  &19.3  &68.2\\
\hhline{~-----------}
CNN         &10  &59   &2  &13   &25    &31   &3   &38    &15   &19.6  &64.4\\ %&15.8  &3.8   &65.4\\
BBC         &6   &75   &2  &50   &98    &96   &1   &21    &0    &29    &71\\   %&29    &0     &71\\
NewsBreak   &27  &66   &5  &25   &152   &92   &5   &63    &0    &20    &80\\   %8     &12    &80\\
\hhline{~-----------}
Gmaps       &17  &65   &6  &46   &870   &99   &4   &57    &0    &37    &63\\   %12    &25    &63\\
Uber        &13  &59   &6  &43   &238    &95   &13   &53    &0    &25    &75\\   %&8     &20    &72\\
\hhline{~-----------}
Letgo       &10  &56   &3  &30   &715   &97   &1   &18    &6    &31    &63\\   %&6     &25    &63\\
Amazon      &33  &67   &5  &45   &1490  &96   &7   &84    &0.4  &30.6  &69\\   %&7.8   &22.8  &69\\
\hhline{~-----------}
Gmail       &1   &14   &1  &20   &16    &91   &2   &82    &0    &46.5  &53.5\\ %&17.6  &28.9  &53.5\\
Outlook     &4   &57   &2  &50   &20    &91   &2   &79    &3    &32    &65\\   %&10    &22    &65\\
\hhline{~-----------}
Youtube     &10  &63   &4  &36   &127   &84   &3   &46    &5.8  &25    &69.2\\ %&9     &16    &69.2\\
SoundCloud  &10  &43   &2  &20   &715   &99   &8   &76    &0    &84    &78\\   %&6     &16    &78\\
Spotify     &1   &13   &2  &25   &78    &95   &5   &59    &1    &15    &84\\   %&9     &6     &84\\
\hhline{~-----------}
AVERAGE     &10  &56   &3  &38   &523   &89   &5   &63    &2.5  &24.8  &72.7\\ %&9.7   &15.1  &72.7\\
\hhline{~===========}
Browsing      &12   &48 &5   &37  &488   &71    &5.53     &38    &2.6    &21.5  &76\\ %&5.5    &16       &76\\
\hhline{~-----------}
\end{tabular}
\hspace{-1pt}
\begin{tabular}{
    |@{\hspace{0pt}} r @{\hspace{1pt}} r @{\hspace{1pt}}
    |@{\hspace{0pt}} r @{\hspace{1pt}} r @{\hspace{1pt}}
    |@{\hspace{0pt}} r @{\hspace{1pt}} r @{\hspace{1pt}}
    |@{\hspace{0pt}} r @{\hspace{1pt}} r @{\hspace{1pt}}
    |@{\hspace{0pt}} r 
     @{\hspace{1pt}} r
     @{\hspace{1pt}} r
     @{\hspace{1pt}}|
    }
\multicolumn{11}{c}{App-click} 
\\
\hhline{-----------}
\multicolumn{2}{|@{\hspace{0pt}}c@{\hspace{1pt}}}{fl.} &
\multicolumn{2}{|@{\hspace{0pt}}c@{\hspace{1pt}}}{dom.} &
\multicolumn{2}{|@{\hspace{0pt}}c@{\hspace{1pt}}}{vol.[kB]} &
\multicolumn{2}{|@{\hspace{0pt}}c@{\hspace{1pt}}}{TC[s]} &
\multicolumn{3}{|@{\hspace{0pt}}c@{\hspace{1pt}}|}{TC break [\%]}
\\
%App         & all   & CP    & all  & CP              \%Dns & \%Tcp & \%Tls/QUIC & \%Data \\ \hline
%Letgo       & 736   & 715   & 3.6  & 0.65            & 6     & 6     & 25         & 63     \\ \hline
%Amazon      & 1560  & 1490  & 8.22 & 6.98            & 0.4   & 7.8   & 22.8       & 69     \\ \hline
%CNN         & 80    & 25    & 7.06 & 2.69            & 15    & 15.8  & 3.8        & 65     \\ \hline
%Gmaps       & 880   & 870   & 7.3  & 4.2              & 0     & 12    & 25         & 63     \\ \hline
%Youtube     & 151   & 127   & 6.59 & 3.03            & 5.8   & 9     & 16         & 68.7   \\ \hline
%Outlook     & 22    & 20    & 2.7  & 2.15             & 3     & 10    & 22         & 65     \\ \hline
%NewsBreak   & 165   & 152   & 7.1  & 4.5              & 0     & 8     & 12         & 80     \\ \hline
%SoundCloud  & 726   & 715   & 10.6 & 8.15            & 0     & 6     & 16         & 78     \\ \hline
%Spotify     & 82    & 78    & 7.8  & 4.6              & 22    & 9     & 6          & 62     \\ \hline
%Uber        & 245   & 61    & 8.3  & 5.36            & 0     & 8     & 20         & 72     \\ \hline
%BBC         & 102   & 98    & 2.46 & 0.52            & 0     & 29    & 0          & 71     \\ \hline
 abs &\%   &abs  &\%   &  abs   &\%    &abs  &\%    & dns   & hshake & data \\
\hhline{-----------}
1   &13   &1  &25   &29    &96   &19  &44   &0   &0   &100\\    %&0   &0        &100 \\
3   &60   &2  &40   &1313  &63   &5   &54   &6   &10  &84\\     %&10  &0.1      &84 \\
4   &57   &1  &50   &1538  &90   &2   &50   &0   &0   &100\\    %&0   &0        &100 \\
\hhline{-----------}
1   &100  &1  &100  &1     &100  &1   &100  &0   &0   &100\\    %&0   &0        &100 \\
2   &22   &1  &33   &194   &75   &6   &17   &2   &7   &91\\     %&4  &3        &91 \\
2   &40   &2  &67   &20    &79   &10  &35   &0   &2   &98\\     %&1  &1       &98 \\
\hhline{-----------}                                            
3   &25   &2  &40   &69    &82   &1   &55   &0   &3   &97\\     %&3   &0        &97 \\
4   &36   &2  &67   &105   &92   &1   &67   &0   &22  &78\\     %&22  &0        &78 \\
19  &43   &7  &78   &96    &20   &3   &73   &0   &13  &87\\     %&7   &6        &87 \\
\hhline{-----------}                                            
3   &60   &2  &100  &870   &98   &2   &52   &0   &0   &100\\    %&0   &0        &100 \\
3   &43   &1  &50   &13    &11   &5   &85   &0   &0   &100\\    %&0   &0        &100 \\
\hhline{-----------}                                            
5   &100  &2  &100  &65    &100  &2   &100  &0   &6   &94\\     %&1   &5        &94 \\
21  &34   &4  &36   &1650  &92   &12  &80   &0   &6   &94\\     %&3   &3        &94 \\
\hhline{-----------}                                            
6   &55   &2  &40   &38    &75   &11  &21   &0   &7   &93\\     %&3  &4       &93 \\
5   &83   &3  &75   &9     &100  &0   &35   &0   &75  &25\\     %&21  &54       &25 \\
\hhline{-----------}                                            
5   &45   &1  &20   &65    &47   &1   &31   &0   &67  &35\\     %&23  &42       &35 \\
1   &17   &1  &33   &120   &99   &1   &44   &0   &0   &100\\    %&0   &0        &100 \\
3   &30   &1  &50   &115   &98   &0   &52   &0   &0   &100\\    %&0   &0        &100\\
\hhline{-----------}                                            
5   &48   &2  &56   &351   &79   &4   &55   &0   &12  &88\\     %&5   &7       &88\\
\hhline{-----------}                                            
\multicolumn{11}{c}{\rule{0pt}{1.2em}}
\end{tabular}

\end{table*}

%App                                #Domains  #Critical_domains  #Flows  #Critical_flows
%com.abtnprojects.ambatana          &5       &5 &2         &2                  
%com.amazon.mShop.android.shopping  &61      &21&11        &4                  
%com.cnn.mobile.android.phone       &12      &3 &5         &2                  
%com.google.android.apps.maps       &5       &3 &2         &2                  
%com.google.android.youtube         &11      &5 &5         &1                  
%com.microsoft.office.outlook       &6       &5 &4         &3                  
%com.particlenews.newsbreak         &44      &19&9         &7                  
%com.soundcloud.android             &6       &1 &3         &1                  
%com.spotify.music                  &10      &3 &2         &1                  
%com.ubercab                        &7       &3 &2         &1                  
%bbc.mobile.news.uk                 &11      &4 &3         &2                  

\subsection{Critical traffic properties across apps} 
Table~\ref{tab:critical} summarizes the critical traffic properties for both \appstartup (left) and \appclick (right). For each app we report the number of critical flows, domains, bytes both in absolute and percentage averaged across different runs. % with respect to the whole traffic. 
We also report the time spent on the critical path (TC) %(i.e., how much of the TDT relates to flows on the critical path), 
and how this is spent doing DNS, transport handshakes, and data transfers. Table rows are grouped by app categories. %, and the last row reports metrics averages across apps, although browsing is not accounted in the aggregate. 

%The goal is not to provide a longitudinal study on critical path of mobile applications, but rather to highlight the importance of critical path analysis and provide preliminary insights from the set of considered applications.

\noindent\textbf{Traffic volume.}
%Results show that the volume of traffic on the critical path is larger than one would expect. 
On average, 56\% (48\%) of flows, 89\% (79\%) of bytes are critical in \appstartup (\appclick). 
Differently from what we expected, in absolute scale the volume of bytes is still significant in \appclick (351kB on average, almost 70\% of the average volume in \appstartup). Considering domains, 38\% are critical in \appstartup startup against 56\% for \appclick. There are macroscopic differences between apps, but no visible patterns within and between categories or scenarios. For instance, Whatsapp is an ``outlier'' as all traffic is carried over 1-2 flows, hence everything is critical. 
The only class that seems different is web browsing, which presents 48\% (71\%) of critical flows (bytes), -8\% (-18\%) with respect to apps startup.
%Amazon presents similar relative values between \appstartup and \appclick, while for Letgo traffic is critical only in \appclick.

\noindent\textbf{Time on CP.} For browsing also TC is lower, 38\% against 63\% (55\%) in \appstartup (\appclick). %As browsing and apps present a similar traffic volume as critical, we conjecture the combination of different effects. 
%On the one hand, browsing forces content to be downloaded even after AFT, hence TDT is longer (\cref{sec:waterfall}).  (4-5s). 
On the other hand, for both browsing and apps TC is similar in absolute scale (4-5s).
In other words, despite the diversity in the actions triggered, results suggest that the differences in the critical traffic between startup and actual app usage could be less pronounced that one might think.
As expected, data transfer has the largest impact on the critical path with 72.7\% (88\%) for \appstartup (\appclick). DNS is generally small except for a few cases.
Conversely, protocol handshakes are heavier at startup (24.8\% on average), but \appclick shows unexpected bi-modal behaviour with either a heavy (\eg 67\% YouTube, 10\% Facebook) or negligible weight. %By investigating on this, we find that many of the flows contributing to the overhead are due to ads or 3rd party services.

\noindent\textbf{Content type analysis.}
%The role of the domains in the traffic is also relevant. In particular, 
Extracting keywords from the domains, we split the traffic in 3 classes: \emph{ad-hoc} (apps/websites specific domains), \emph{cdn}, and \emph{oth-serv} (\eg 3rd party services, ad networks).
We find that for apps (browsing) TC is split into 68\% (33\%), 25\% (51\%), and 7\% (15\%), while volume is split into 47\% (25\%), 52\% (65\%), and 1\% (9\%) for ad-hoc, cdn, and oth-serv respectively.
In other words, apps network latency tends to gravitate towards app-specific domains. Those are not necessarily responsible only for control logic as they carry almost the same volume as CDNs. Conversely, browsing content is likely served by CDNs. 
Considering oth-serv, browsing spends 2$\times$ TC than apps, but downloads 9$\times$ more volume than apps.

\begin{figure}[!t]
\includegraphics[width=\columnwidth]{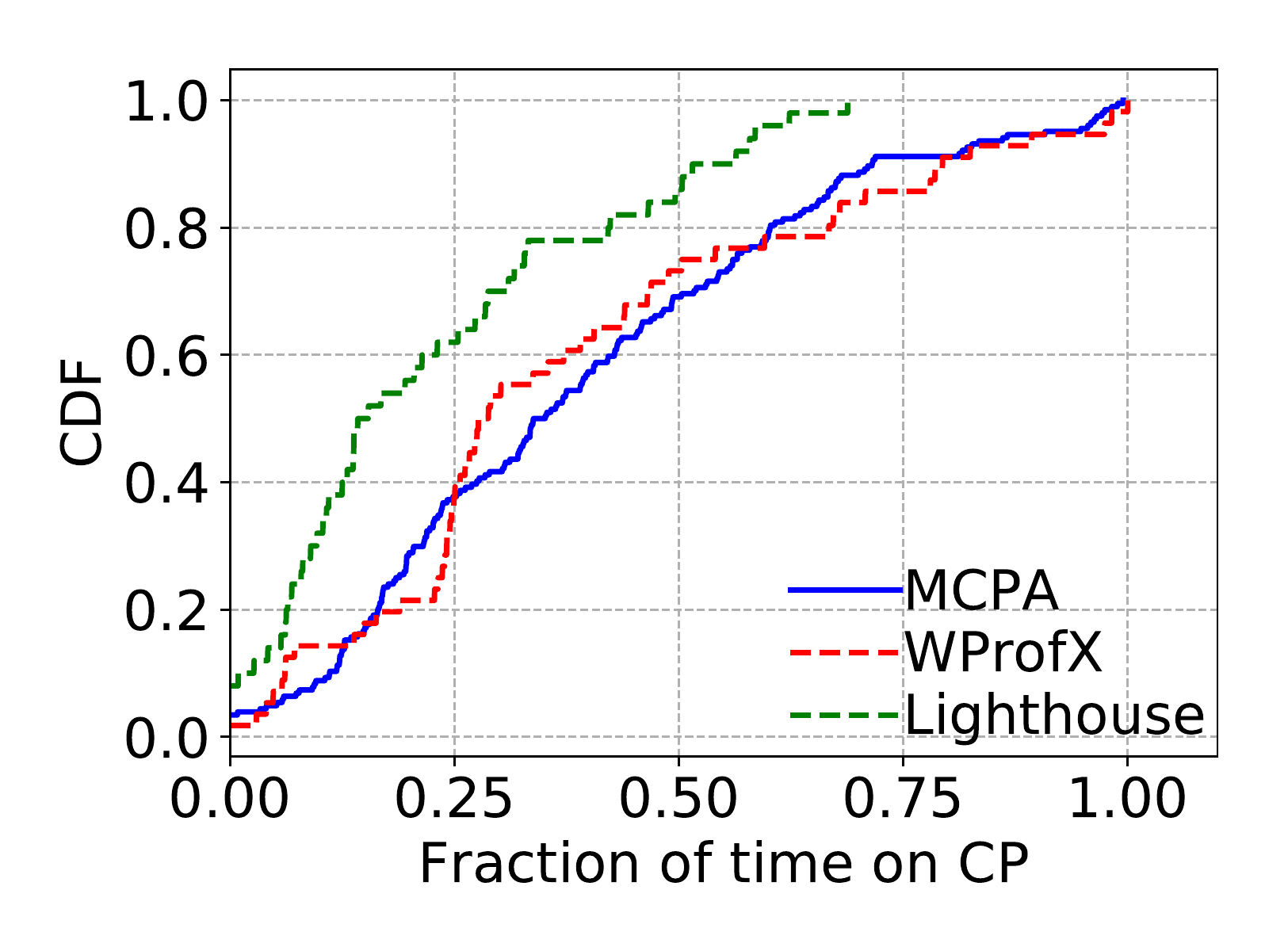}
\caption{Comparing network time on CP across CPA tools.
\label{fig:mcpa-lighthouse-domains}}
\end{figure}

\begin{figure}[!t]
\centering
\includegraphics[width=0.45\textwidth]{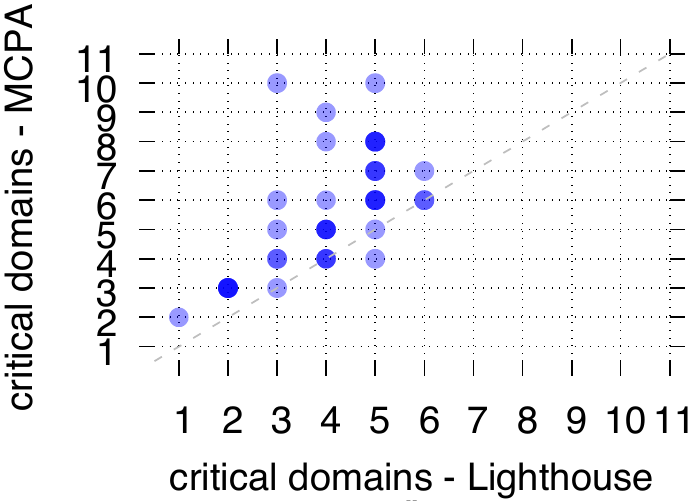}
\caption{Comparing MCPA and Lighthouse: number of critical domains. % (left); time spent on critical path (right)
\label{fig:mcpa-vs-lighthouse}}
\end{figure}

\begin{figure}[!t]
\centering
\includegraphics[width=0.45\textwidth]{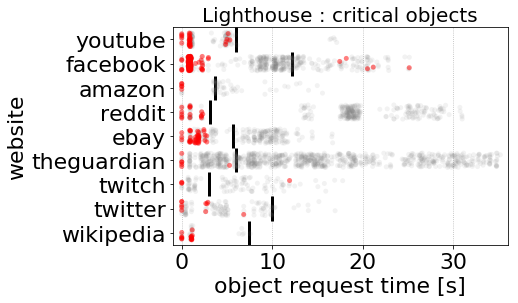}
\hspace{-7pt}
\includegraphics[width=0.45\textwidth]{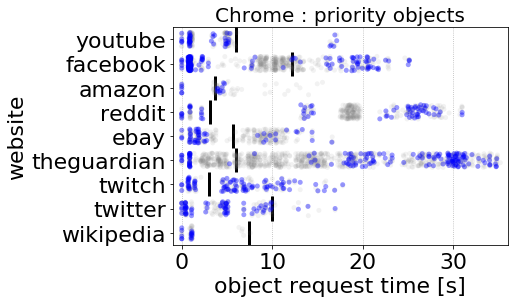}
\caption{{Lighthouse critical path analysis.} % All requests in grey, priority objects in blue and critical objects in red.}
\label{fig:lighthouse-strip}}
\end{figure}

\section{Discussion}
\label{sec:discussion}
\tool aims to identify critical traffic generated by generic mobile apps. A few other CPA tools for mobile apps have been presented, but none of them are applicable to our intent as they either require heavy on-device instrumentation or do not dissect traffic dynamics~\cite{appinsightOSDI12,panappticonISSS13}. 
%Instead, focusing on web traffic only, Google Lighthouse~\cite{lighthouse} seems an interesting tool. It is both open sourced, and integrated into Chrome's devtools suite, and offers a fine-grained auditing service for webpages which includes \emph{Critical Request Chains} (CRCs) pinpointing objects generating bottlenecks.\footnote{https://developers.google.com/web/tools/lighthouse/audits/critical-request-chains}
However, restricting the focus to web browsing only, we can compare \tool with WProfX (the WProf version for mobile browsing) and Google Lighthouse, both open sourced.
Fig.~\ref{fig:mcpa-lighthouse-domains} shows the CDF of the fraction of time on CP for the three tools. We highlight that for \tool and Lighthouse, time on CP implicitly refers to only network activities, while WProfX reports also on parsing and rendering time, which we exclude for the comparison.

%We initially consider WProf, which has been widely adopted in the literature. 
\noindent\textbf{WProfX} profiles the impact of webpages loading activities on PLT. %, and broadly classify them as either \emph{network} or \emph{computation}. %As such, we can quantify the impact of WProf critical traffic on the PLT\footnote{As to the PLT, we consider the time the Load event is triggered}, and compare it with \tool critical traffic.  As depicted in  
%Fig.~\ref{fig:mcpa-vs-wprof} compares the CDF of the fraction of time on CP with for both WProfX and \tool. 
Notice the strong similarity of \tool and WProfX CDFs, with both tools reporting 38\% of time on CP on average. This implies that \tool, even if based on traffic analysis only, is comparable with an in-browser profiling engine.
%, the results of \tool are well aligned with WProf, with the critical traffic accounting for the 38\% of the page load time on average in both cases. 
%More interestingly, such percentage is far below the results obtained for the mobile apps, as shown in Fig.~\ref{fig:mcpa-vs-wprof}(right), where the critical traffic consumes on average the 55\% of TDT for \appclick and the 63\% for \appstartup. This suggests that more "specialized" app software (and multi-threading) imply a much stronger impact of the network on the overall latency. 
%\gtnote{this needs some rephrase}.     

\noindent\textbf{Lighthouse} %Finally, we consider the case of Lighthouse. Among its services, Google's tool 
reports the webpage \emph{Critical Request Chains} (CRCs) pinpointing to objects generating bottlenecks.\footnote{\tiny \url{https://developers.google.com/web/tools/lighthouse/audits/critical-request-chains}}
As visible in Fig.~\ref{fig:mcpa-lighthouse-domains}, Lighthouse reports a shorter time on CP than both WProfX and \tool. We found that \tool generally classifies a few more domains as critical than Lighthouse (Fig.~\ref{fig:mcpa-vs-lighthouse}), but the same is true for WProfX too.
%When comparing MCPA results with Lighthouse output, The left scatter plot shows the number of critical domains, while the right one compares the time spent on CP as seen by the tools. \tool generally finds a few more critical domains than Lighthouse, and the time spent on the critical path is longer as well. Inspecting those differences, we find that in most cases they are due to Lighthouse internal mechanisms. We highlight that Lighthouse logic is not detailed by the online doc, so we investigated the source code. 
The reason of the discrepancy resulted clear only by investigating Lighthouse source code, \ie it is due to an internal design choice not publicly documented.
Specifically, Lighthouse marks objects as critical if they have a \emph{network priority} higher than medium (\ie the browser schedules objects fetch early on), and they are neither images, XML HTTP Request (XHR), nor server push(ed) content. This results in a ``constrained'' view of the traffic as reported for a subset of websites by the strip-plots in Fig.~\ref{fig:lighthouse-strip}: grey dots represent all requests; red dots (left plot) mark critical objects; blue dots (right plot) marks prioritized objects; vertical black lines mark the AFT. Notice how Lighthouse is biased towards the first part of the download, which possibly involves only ``structural'' properties of the webpage rather than actual content.

%\section{Conclusion and Discussion}
%\label{sec:conclusion}
Beside the fine-grained details, the tools comparison highlights a more subtle problem: the lack of standard methodologies to pinpoint what is critical, and how to perform root cause analysis related to those bottlenecks. These goals go beyond the purpose of our work, which instead addresses a prior and more fundamental requirement: to ease the study of generic mobile apps. We demonstrated that network measurements can be effective and easier to adopt than rendering based metrics such as AFT/SI. Moreover, our definition of critical path aims to discover any critical network activity without any restriction on the type, so to capture traffic dynamics as a whole.
To test \tool we adopted the standard practice of an instrumented device, with the intention to demonstrate that this might not be necessary. 
%Beside the fine-grained details, the tools comparison highlights a more subtle problem: the lack of standard way to pinpoint what is critical, and how to perform root cause analysis related to those bottlenecks. Neither of the considered tools, nor any of the CPA tool developed so far can solve the issue, because none consider the human feedback to understand what really is critical. Such an effort however goes beyond the purpose of our work, which was centered around a prior and fundamentail requirement: ease the way to study mobile apps. 
%In particular, in this work we validated the idea of using network measurements to create proxy metrics of AFT and SI. 
This can open the doors to a new class of tools easier to deploy than current state of the art techniques, without significantly sacrificing accuracy. 
In this way, app developers and mobile operators could better dissect traffic dynamics (\eg TCP/TLS handshake, TCP fast open~\cite{tcpfastCONEXT11}, app-specific protocols, control logic, or pre-fetching) by means of at-scale measurement campaigns.

\begin{table*}[!t]
\footnotesize
\caption{App usage patterns.\label{tab:usagepatterns}}
\begin{tabular}{r l p{30em} p{0em}}
\bf app name & \bf package name & usage pattern \\
\hhline{====}
Twitter         &com.twitter.android            &
    start app; open {\it trending now}; select top result; select top tweet; refresh home. 
    \\
\hline
Facebook        &com.facebook.katana            &
    start app; move to {\it notifications}; select top notification; move back to home; refresh home. 
    \\
\hline
Instagram       &com.instagram.android          &
    start app; open {\it search} tab; move back to home (refresh); open {\it story} from top bar.  
    \\
\hline
Whatsapp& com.whatsapp & 
    start app; select top conversation; type random message and send (x3).
    \\
\hline
Snapchat        &com.snapchat.android           &
    start app; select friend from list; grab and send picture (x3). 
    \\
\hline
Messenger       &com.facebook.orca critical     &
   start app; open top conversation; type random message and send (x3).
    \\
\hline
CNN             &com.cnn.mobile.android.phone   &
    start app; open top news; open next news; move back to homepage; open CNN video portal
    \\
\hline
BBC             &bbc.mobile.news.uk             &
    start app; open top news; move to popular news list; move back to homepage; open {\it My News} area.
    \\
\hline
NewsBreak       &com.particlenews.newsbreak     &
    start app; move to news category (\eg World, Business); open top result; move to next category; open top result.     
     \\
\hline
Gmaps           &com.google.android.apps.maps   &
    start app; open tab {\it explore restaurants}; select top result; tap on {\it indications}; show route info. 
     \\
\hline
Uber            &com.ubercab                    &
    start app; open search box; select destination from history; cancel; back to homepage.
    \\
\hline
Letgo           &com.abtnprojects.ambatana      &
    start app; open random item category (\eg tech); open top offer; tap on {\it more info} to display item details (including geographical location);  move to different category.
    \\
\hline
Amazon          &com.amazon.mShop.android.shopping &
   start app; open side menu; select top offers; select top item; open item details page.
    \\
\hline
Gmail           &com.google.android.gm          &
   start app; open random email from inbox; open {\it reply} tab; send empty reply; move back to inbox.
       \\
\hline
Outlook         &com.microsoft.office.outlook   &
   start app; open random email from inbox; open {\it reply} tab; send empty reply; move back to inbox.
    \\
\hline
Youtube         &com.google.android.youtube     &
    start app; move to {\it trending} tab; move to {\it subscriptions} tab;  select top result (playback starts); exit playback. 
    \\
\hline
SoundCloud      &com.soundcloud.android         &
    start app; open {\it liked tracks} tab; open top song; start song playback; exit playback.
    \\
    \hline
Spotify         &com.spotify.music              &
    start app; move to {\it your playlists}; open top playlist; start playback; stop playback (and move back to home).
    \\
\hline
\end{tabular}
\end{table*}

\begin{table*}[!t]
\footnotesize
\caption{Critical domains.\label{tab:criticaldomains}}
\begin{tabular}{r l p{15em} p{15em}}
\bf name & \bf package name & \appclick & \appstartup \\
\hhline{====}
Twitter         &com.twitter.android            &
    api.twitter.com 
    &
    api.twitter.com\\
\hline
Facebook        &com.facebook.katana            &
    graph.facebook.com\newline 
    external-lht6-1.xx.fbcdn.net
    &
    graph.facebook.com\newline
    scontent-lhr3-1.xx.fbcdn.net\newline
    external-lhr3-1.xx.fbcdn.net
    \\
\hline
Instagram       &com.instagram.android          &
    scontent-lht6-1.cdninstagram.com 
    &
    scontent-lhr3-1.cdninstagram.com\newline
    i.instagram.com
    \\
\hline
Whatsapp&& &\\
\hline
Snapchat        &com.snapchat.android           &
    mvm.snapchat.com
    &
    app.snapchat.com\newline
    app-analytics.snapchat.com\newline
    mvm.snapchat.com\newline
    sc-analytics.appspot.com
    \\
\hline
Messenger       &com.facebook.orca critical     &
    edge-mqtt.facebook.com\newline 
    lookaside.facebook.com
    &
    b-graph.facebook.com\newline
    scontent.xx.fbcdn.net
    \\
\hline
CNN             &com.cnn.mobile.android.phone   &
    cerebro.api.cnn.io\newline
    dynaimage.cdn.turner.com
    &
    cerebro.api.cnn.io\newline
    dynaimage.cdn.turner.com
    \\
\hline
BBC             &bbc.mobile.news.uk             &
    trevor-producer-cdn.api.bbci.co.uk\newline 
    ichef.bbci.co.uk
    &
    trevor-producer-cdn.api.bbci.co.uk\newline
    ichef.bbci.co.uk
    \\
\hline
NewsBreak       &com.particlenews.newsbreak     &
    graph.facebook.com\newline
    googleads.g.doubleclick.net\newline 
    lh3.googleusercontent.com\newline
    api.particlenews.com\newline
    api.mobula.sdk.duapps.com\newline
    img.particlenews.com\newline
    log.particlenews.com
    &
    graph.facebook.com\newline
    googleads.g.doubleclick.net\newline
    static.xx.fbcdn.net\newline
    tpc.googlesyndication.com\newline
    scontent.xx.fbcdn.net
    \\
\hline
Gmaps           &com.google.android.apps.maps   &
    clients4.google.com\newline
    www.google.com
    &
    www.google.com\newline
    lh4.googleusercontent.com\newline
    lh3.googleusercontent.com\newline
    lh5.googleusercontent.com\newline
    lh6.googleusercontent.com\newline
    clients4.google.com
    \\
\hline
Uber            &com.ubercab                    &
    cn-sjc1.uber.com
    &
    csi.gstatic.com\newline
    d1w2poirtb3as9.cloudfront.net\newline
    api.braintreegateway.com\newline
    cn-sjc1.uber.com\newline
    cn-geo1.uber.com\newline
    clients4.google.com
    \\
\hline
Letgo           &com.abtnprojects.ambatana      &
    search-products.letgo.com\newline
    img.letgo.com
    &
    stickers.letgo.com\newline
    search-products.letgo.com\newline
    img.letgo.com\\
\hline
Amazon          &com.amazon.mShop.android.shopping &
    m.media-amazon.com\newline
    www.amazon.es\newline
    images-eu.ssl-images-amazon.com\newline
    images-na.ssl-images-amazon.com
    &
    m.media-amazon.com\newline
    msh.amazon.co.uk\newline
    images-eu.ssl-images-amazon.com\newline
    images-na.ssl-images-amazon.com\newline
    www.amazon.es
    \\
\hline
Gmail           &com.google.android.gm          &
    ci5.googleusercontent.com\newline
    inbox.google.com
    &
    inbox.google.com
    \\
\hline
Outlook         &com.microsoft.office.outlook   &
    prod15-api.acompli.net\newline
    prod15-files.acompli.net\newline
    mobile.pipe.aria.microsoft.com
    &
    prod15-api.acompli.net\newline
    prod15-files.acompli.net
    \\
\hline
Youtube         &com.google.android.youtube     &
    s.ytimg.com
    &
    i.ytimg.com\newline
    www.youtube.com\newline
    youtubei.googleapis.com\newline
    yt3.ggpht.com
    \\
\hline
SoundCloud      &com.soundcloud.android         &
    i1.sndcdn.com
    &
    api-mobile.soundcloud.com\newline
    i1.sndcdn.com
    \\
    \hline
Spotify         &com.spotify.music              &
    i.scdn.co
    &
    i.scdn.co\newline
    pl.scdn.co
    \\
\hline
\end{tabular}
\end{table*}

%\clearpage
%\bibliographystyle{splncs04}
\bibliographystyle{ACM-Reference-Format}
\bibliography{biblio}

\end{document}